\pdfoutput=1
\RequirePackage{ifpdf}
\ifpdf % We are running pdfTeX in pdf mode
\documentclass[pdftex]{sigma}
\else
\documentclass{sigma}
\fi

\numberwithin{equation}{section}

\newcommand{\so}{\scriptscriptstyle \rm I}
\newcommand{\st}{\scriptscriptstyle \rm I\hspace{-1pt}I}

\newtheorem{thm}{Theorem}[section]
\newtheorem{cor}[thm]{Corollary}
{ \theoremstyle{definition}
\newtheorem{Rem}[thm]{Remark}
}

\begin{document}

\newcommand{\arXivNumber}{1501.07566}

\allowdisplaybreaks

\renewcommand{\PaperNumber}{063}

\FirstPageHeading

\ShortArticleName{${\rm GL}(3)$-Based Quantum Integrable Composite Models. I.~Bethe Vectors}

\ArticleName{$\boldsymbol{{\rm GL}(3)}$-Based Quantum Integrable Composite Models. I.~Bethe Vectors}

\Author{Stanislav~PAKULIAK~${}^{abc}$, Eric~RAGOUCY~${}^d$ and Nikita~A.~SLAVNOV~${}^e$}

\AuthorNameForHeading{S.~Pakuliak, E.~Ragoucy and N.A.~Slavnov}

\Address{$^a$~Institute of Theoretical \& Experimental Physics, 117259 Moscow, Russia}

\Address{$^b$~Laboratory of Theoretical Physics, JINR, 141980 Dubna, Moscow reg., Russia}
\EmailD{\href{mailto:pakuliak@jinr.ru}{pakuliak@jinr.ru}}

\Address{$^c$~Moscow Institute of Physics and Technology, 141700, Dolgoprudny, Moscow reg., Russia}

\Address{$^d$~Laboratoire de Physique Th\'eorique LAPTH,
CNRS and Universit\'e de Savoie, \\
\hphantom{$^d$}~BP 110, 74941 Annecy-le-Vieux Cedex, France}
\EmailD{\href{mailto:eric.ragoucy@lapth.cnrs.fr}{eric.ragoucy@lapth.cnrs.fr}}

\Address{${}^e$ Steklov Mathematical Institute, Moscow, Russia}
\EmailD{\href{mailto:nslavnov@mi.ras.ru}{nslavnov@mi.ras.ru}}

\ArticleDates{Received February 18, 2015, in f\/inal form July 22, 2015; Published online July 31, 2015}

\Abstract{We consider a composite generalized quantum integrable model solvable by the nested algebraic
Bethe ansatz.
Using explicit formulas of the action of the monodromy matrix elements onto Bethe vectors in the
${\rm GL}(3)$-based quantum integrable models we prove a formula for the Bethe vectors of
composite  model. We show that this representation is a~particular case of general coproduct property
of the weight functions (Bethe vectors) found in the theory
of the deformed Knizhnik--Zamolodchikov equation.}

\Keywords{Bethe ansatz; quantum af\/f\/ine algebras, composite models}

\Classification{17B37; 81R50}

\section{Introduction}

Solution of the inverse scattering problem for XXZ Heisenberg spin-$1/2$
chain  in the pioneering  paper \cite{KMT99} opened a possibility to apply the results of calculations of the Bethe vectors
scalar products  \cite{Sl89} to the problem of calculation of the correlation functions
for some class of quantum integrable models (see review \cite{Sl07} and references therein).
This class of integrable models is characterized by the property that the monodromy operator is constructed
from the $R$-matrix and  it becomes
a permutation operator at some value of the spectral parameter.

For quantum models where such property is absent there exists another method to calculate the correlation
functions of the local operators \cite{IzK84}. The
main idea of this method is to divide artif\/icially the interval where model is def\/ined into two  sub-intervals.
In this framework one has a possibility to study form factors of operators depending on an internal point of the original interval.
This approach was used  to analyze some correlation functions in the  bose gas model that describes one-dimensional bosons with $\delta$-function
interaction \cite{IzKR87,KitKMST07}. The authors of \cite{IzK84} called such the model {\it two-site}
generalized model. Later, in \cite{Sl07}  the terminology {\it two-component} model was used. We think that both
terms might be misleading: the f\/irst one if we apply this approach to the study of lattice models (like spin chains,  where `site' has already a def\/inite and dif\/ferent meaning);
the second if we use such model for the description of multi-component gases,  where `component' refers to the dif\/ferent types of
particles. Therefore
in this paper we call this model a {\it composite} model.

In the   composite model the  total monodromy matrix $T(u)$  is
presented as a usual matrix product of the partial monodromy matrices $T^{(2)}(u)$ and $T^{(1)}(u)$:
\begin{gather}\label{T-TT}
T(u)=T^{(2)}(u)T^{(1)}(u).
\end{gather}
The matrix elements of $T(u)$ are operators in the space of states $V$ that corresponds to an interval $[0,L]$.
The matrix elements of the partial monodromy matrices $T^{(1)}(u)$ and $T^{(2)}(u)$  act in the
spaces $V^{(1)}$ and $V^{(2)}$  corresponding
to the intervals $[0,x]$ and $]x,L]$ respectively. Here $x$ is an intermediate point of the interval $[0,L]$.
The total space of states $V$ is a tensor product of the partial spaces of states $V^{(1)}\otimes V^{(2)}$.
We assume that each of these spaces possesses unique vacuum vector  $|0\rangle^{(\ell)}$ def\/ined by the formulas
\begin{alignat}{3}
& T_{ij}(u)|0\rangle =0,\quad i>j,\qquad && T_{ii}(u)|0\rangle=\lambda_i(u)|0\rangle,& \nonumber\\
& T^{(\ell)}_{ij}(u)|0\rangle^{(\ell)} =0,\quad i>j,\qquad && T^{(\ell)}_{ii}(u)|0\rangle^{(\ell)}=\lambda^{(\ell)}_i(u)|0\rangle^{(\ell)},\quad \ell=1,2,& \label{vac-vec}
\end{alignat}
and $|0\rangle=|0\rangle^{(1)}\otimes |0\rangle^{(2)}$.

We also assume that dual spaces ${V^*}^{(1)}$ and ${V^*}^{(2)}$ possess a dual vacuum vectors $\langle0|^{(\ell)}$ with analogous properties
\begin{alignat*}{3}
& \langle0|T_{ij}(u) =0,\quad i<j,\qquad &&  \langle0|T_{ii}(u)=\lambda_i(u)\langle0|,& \\
& \langle0|^{(\ell)} T^{(\ell)}_{ij}(u) =0,\quad i<j,\qquad && \langle0|^{(\ell)}T^{(\ell)}_{ii}(u)=\lambda^{(\ell)}_i(u)\langle0|^{(\ell)},\quad \ell=1,2,& %\label{dvac-dvec}
\end{alignat*}
and $\langle0|=\langle0|^{(1)}\otimes \langle0|^{(2)}$.

Spaces of states $V^{(1)}$, $V^{(2)}$ in each part of the composite  model are formed by  partial Bethe vectors $\mathbb{B}^{(\ell)}\in
V^{(\ell)}$, $\ell=1,2$. They are given by certain polynomials in the entries of the partial monodromy matrices
$T^{(\ell)}_{ij}(u)$, $i<j$ acting onto  vacuum vector $|0\rangle^{(\ell)}$.  Total Bethe vectors $\mathbb{B}\in V$ are the same polynomials in  the entries of the total monodromy matrix acting onto~$|0\rangle$. Bethe vectors (total and partial) depend
on  complex variables
called Bethe parameters (see, e.g.,~\eqref{BV-explicit}). Total Bethe vectors are
characterized by the property that at certain values of Bethe parameters they become eigenvectors of a total transfer matrix~$\operatorname{tr} T(u)$, where the trace
is taken in the matrix space. Similarly,
partial Bethe vectors are eigenvectors of  partial transfer matrices $\operatorname{tr} T^{(\ell)}(u)$ at certain values of their Bethe parameters.

The main advantage of the composite model is that it allows one to calculate form factors of the partial monodromy matrix elements
$T^{(\ell)}_{ij}(u)$ in the basis of the total Bethe vectors. This opens a way for computing form factors and correlation functions
of local operators~\cite{IzK84,IzKR87,KitKMST07}. However, for this purpose, one should express the total Bethe vectors in terms of the
partial ones. Such  representation is a necessary tool for all further studies.

This problem was solved  for ${\rm GL}(2)$-based models in~\cite{IzK84}. There it was shown that the total Bethe vector is a linear combination
of the partial Bethe vectors tensor products. This is not a~surprising result, because in the ${\rm GL}(2)$-based models Bethe vectors have a very simple structure:
they are monomials in the operator~$T_{12}(u)$ acting on~$|0\rangle$. Since due to~\eqref{T-TT} we
have
\begin{gather}\label{TikTkj}
T_{ij}(u)=\big( T^{(2)}(u)\cdot T^{(1)}(u)\big)_{ij} =\sum_{k=1}^2T_{ik}^{(2)}(u)T_{kj}^{(1)}(u),
\end{gather}
we f\/ind, in particular, $T_{12}(u)=T_{11}^{(2)}(u)T_{12}^{(1)}(u)+T_{12}^{(2)}(u)T_{22}^{(1)}(u)$. It is clear therefore that the action of any monomial
in $T_{12}(u)$ on the vacuum reduces to a bilinear combination of monomials in $T_{12}^{(\ell)}(u)$ acting on partial
vacuums\footnote{Note that in~\eqref{TikTkj}
the operators $T_{ik}^{(2)}(u)$ and $ T_{kj}^{(1)}(v)$ commute with each other, as they act in dif\/ferent spaces.}.
One should only f\/ind the coef\/f\/icients of this bilinear combination. It was done in~\cite{IzK84} (see~\eqref{BV-BVa0} below).

The analogous problem  for ${\rm GL}(N)$-based models with $N>2$ is more sophisticated. In these models Bethe vectors have much more complex structure (see, e.g., \eqref{BV-explicit} for $N=3$). Therefore, the possibility to express total Bethe vectors in terms of partial ones does not look  obviously solvable.

This problem was studied and solved in the theory of the deformed Knizhnik--Zamolodchi\-kov (KZ) equations \cite{TV95}.
In this theory one of the building blocks to construct the integral solutions to these equations is a so-called {\it weight function}.
The weight function appears to be  nothing but a Bethe vector for the model whose monodromy matrix is constructed from the quantum $R$-matrices corresponding to the deformed KZ equation. The weight function is def\/ined in the f\/inite-dimensional representation space of the corresponding
Yang--Baxter algebra  of the monodromy operators~$T_{ij}(u)$. It   should   satisfy a coproduct property. This property states that if the weight function is known for  two representation spaces~$V^{(1)}$ and~$V^{(2)}$, then it can be uniquely determined  for the representation space
$V^{(1)}\otimes V^{(2)}$  (see~\eqref{weight22}).
 If monodromy operators are composed from the generators of some f\/inite~\cite{TV95} or inf\/inite~\cite{KhP08} algebras with Hopf structures,
the coproduct property of the weight functions is a direct consequence of the structure of the Bethe vectors and
the coproduct property of the  monodromy operators
\begin{gather}\label{uni-copro}
\Delta T_{ij}(u)=\sum_{k=1}^{3} T_{kj}(u)\otimes T_{ik}(u),
\end{gather}
applied to the tensor product $V^{(1)}\otimes V^{(2)}$.
 According to the algebraic approach~\cite{KhP08,TV95} monodromy operator for the composite model is def\/ined by the right hand side of the
coproduct formula~\eqref{uni-copro} and using Sweedler  notations this can be rewritten
\begin{gather}\label{uni-copro1}
\sum_{k=1}^{3}  T^{(2)}_{ik}(u) T^{(1)}_{kj}(u)=\big( T^{(2)}(u)\cdot T^{(1)}(u)\big)_{ij}
\end{gather}
in the form of the matrix multiplication of the partial monodromies of the composite model.
Comparing~\eqref{uni-copro1} with~\eqref{T-TT} we see that the usual matrix product in the auxiliary space
can be interpreted as the coproduct in a quantum space, so that~\eqref{T-TT} can be recast as
$\Delta T(u)=T^{(2)}(u)T^{(1)}(u)$. Thus, one can consider the relationship between total and partial Bethe vectors as a coproduct property of the Bethe vectors.

In \cite{TV95} this property was proved  by use of the trace formula for the Bethe vectors and co\-pro\-duct
formulas~\eqref{uni-copro}. Another approach to this problem was developed in~\cite{BeKhP07}.  There
a~{\it universal} weight function was identif\/ied with certain projections of the product of the~$U_q(\widehat{\mathfrak{gl}}_N)$ currents
onto intersections of the Borel subalgebras of dif\/ferent type in the quantum af\/f\/ine algebras. It is worth mentioning that this method also allows one to obtain explicit formulas for the Bethe vectors in~${\rm GL}(N)$-based
quantum integrable models~\cite{KhPT,KhP08,PRS14} in terms of the polynomials of the matrix elements of the monodromy applied to the vacuum vectors.

In this paper we develop one more approach of f\/inding an expression of total Bethe vectors in terms of partial Bethe vectors in ${\rm GL}(3)$-invariant
composite model. We show that this expression can be found directly from the algebra of the monodromy operators~\eqref{RTT-al}. More precisely,
we use the formulas of the action of the monodromy matrix elements onto the Bethe vectors found in \cite{BPRS-BV-12}. As a result,   we present the total Bethe vectors~$\mathbb{B}$
as a linear combination of the tensor products of the partial Bethe vectors\footnote{In what
follows we do not write the symbol $\otimes$ in the tensor product
of the partial Bethe vectors $\mathbb{B}^{(1)}\mathbb{B}^{(2)}$.  Instead we distinguish the
tensor components by the superscripts $^{(1)}$ and $^{(2)}$.} $\mathbb{B}^{(\ell)}$.
The use of morphisms of the algebra~\eqref{RTT-al} also allows us to obtain explicit expressions for dual Bethe vectors of the composite model. The last ones are necessary for calculating form factors of the partial monodromy matrix entries~$T^{(\ell)}_{ij}(u)$. Finally, we show that our results agree
with the coproduct properties of the universal weight functions \eqref{weight22}.

The paper is organized as follows. In Section~\ref{MT} we formulate the main statement and reduce the proof of this assertion
to the calculation of the action of the monodromy matrix elements~$T_{13}(z)$ and~$T_{12}(z)$ onto certain combination of the partial
Bethe vectors $\mathbb{B}^{(\ell)}$. These actions are considered in Sections~\ref{T13} and~\ref{T12}.
Then in Section~\ref{GLN} we specialize the general co\-pro\-duct properties of the universal weight function in ${\rm GL}(N)$-based
model to the ${\rm GL}(3)$ case.  It demonstrates that the direct calculations by the action formulas and coproduct properties
of the universal weight function are well correlated. Appendix~\ref{Action} gathers formulas for the action of the
monodromy matrix elements onto Bethe vectors. Appendix~\ref{Vanish} collects necessary calculations
 for the statements of Section~\ref{T12}.

\section{Coproduct property of  Bethe vectors}\label{MT}

The quantum integrable models considered in this paper are related to the
quantum ${\rm GL}(3)$-invariant $R$-matrix
\begin{gather}\label{R-mat}
R(x,y)=\mathbf{1}+g(x,y)\mathbb{P},\qquad g(x,y)=\frac{c}{x-y},
\end{gather}
where $\mathbb{P}$ is a permutation operator of two three-dimensional spaces~$\mathbb{C}^3$
and~$c$ is a constant. The total monodromy matrix $T(u)$  satisf\/ies the $RTT$ algebra
\begin{gather}\label{RTT-al}
R(w_1,w_2)\cdot (T(w_1)\otimes 1)\cdot (1\otimes T(w_2))=(1\otimes T(w_2))\cdot (T(w_1)\otimes 1)\cdot R(w_1,w_2).
\end{gather}
 Every partial monodromy matrix $T^{(\ell)}(u)$ also satisf\/ies the
$RTT$-relation \eqref{RTT-al} with $R$-mat\-rix~\eqref{R-mat}.
Matrix elements of the monodromy are introduced by the sum $T(u)=\sum_{i,j=1}^3 {\mathsf{E}}_{ij}T_{ij}(u)$, where ${\mathsf{E}}_{ij}$ are
the matrix units with 1 at the intersection of $i^{\rm th}$ row and $j^{\rm th}$ column and zero elsewhere.
In terms of the matrix units $R$-mat\-rix~\eqref{R-mat}
has the form
 \begin{gather*}%\label{R-mat-unit}
R(x,y)=\sum_{i,j=1}^3{\mathsf{E}}_{ii}\otimes{\mathsf{E}}_{jj}+g(x,y)\sum_{i,j=1}^3{\mathsf{E}}_{ij}\otimes{\mathsf{E}}_{ji}.
\end{gather*}

Instead of the functions $\lambda_i(u)$, $\lambda^{(1)}_i(u)$ and $\lambda^{(2)}_i(u)$ def\/ined by \eqref{vac-vec} we will use their ratios
\begin{gather*}%\label{rk}
 r_{k}(u)=\frac{\lambda_{k}(u)}{\lambda_{2}(u)},\qquad
r_{k}^{(\ell)}(u)=\frac{\lambda_{k}^{(\ell)}(u)}{\lambda_{2}^{(\ell)}(u)}, \qquad \ell=1,2, \quad k=1,3.
\end{gather*}
It is obvious from equation \eqref{T-TT} and def\/inition \eqref{vac-vec} of the vacuum vectors that
\begin{gather}\label{lr}
\lambda_{i}(u)=\lambda_{i}^{(1)}(u)\lambda_{i}^{(2)}(u), \qquad
r_{k}(u)=r_{k}^{(1)}(u)r_{k}^{(2)}(u).
\end{gather}

 In this paper we use the same notation and conventions as in \cite{BPRS-BV-12}.
 Besides the rational  function $g(x,y)$ we also use another rational function
\begin{gather}\label{f-fun}
f(x,y)=1+g(x,y)=\frac{x-y+c}{x-y} .
\end{gather}

We  denote sets of variables by bar: $\bar u$, $\bar v$ and so on. We also consider partitions of
sets into disjoint subsets and denote them by symbol $\Rightarrow$.
For example, the notation $\bar u\Rightarrow\{\bar u_{\so}, \bar u_{\st}\}$ means that the
set $\bar u$ is divided into two disjoint subsets\footnote{To be rigorous, since $\bar u_{\so}$ and $\bar u_{\st}$ are sets, we should note $\bar u_{\so}\cup \bar u_{\st}$, but to lighten presentation when there are many subsets, we use the notation $\{\bar u_{\so}, \bar u_{\st}\}$.}
$\bar u_{\so}$ and $\bar u_{\st}$, such that
$\bar u_{\so}\cap\bar u_{\st}=\varnothing$ and $\{\bar u_{\so},\bar u_{\st}\} =\bar u$.
Special notation $\bar u_{0}$, $\bar v_{k}$ and so on  is used for the sets
$\bar u\setminus u_0$, $\bar v\setminus v_k$ etc.

In order to avoid excessively cumbersome formulas we use shorthand notation for products
of commuting operators $T_{ij}(u)$ or the vacuum eigenvalues $\lambda_i(u)$ or $r_k(u)$ (partial and total). Namely, whenever such an operator (or a function) depends on a set of variables, this means that we
deal with the product of these operators (functions) with respect to the corresponding set, as follows:
 \begin{gather}\label{SH-prod}
 T_{23}(\bar v_\ell)= \prod_{\substack{v_k\in\bar v\\v_k\ne v_\ell}} T_{23}(v_k), \qquad\lambda_i(\bar u)=\prod_{u_j\in\bar u}  \lambda_i(u_j),\qquad
  r_1^{(2)}(\bar u_{\st})=\prod_{u_j\in\bar u_{\st}} r_1^{(2)}(u_j).
 \end{gather}
This notation is also used for  products of the functions $f(x,y)$ and $g(x,y)$,
\begin{gather}\label{exam}
g(w,{\bar v})=\prod_{v_j\in{\bar v}} g(w,v_j),\qquad
f(\bar u_{\st},\bar u_{\so})=\prod_{u_j\in\bar u_{\st}}\prod_{u_k\in\bar u_{\so}} f(u_j,u_k).
\end{gather}

\subsection{Bethe vectors}

Bethe vectors of ${\rm GL}(3)$-invariant models depend on two sets of Bethe parameters: $\mathbb{B}=
\mathbb{B}_{a,b}(\bar u;\bar v)$, where ${\bar u}=\{u_1,\ldots,u_a\}$, ${\bar v}=\{v_1,\ldots,v_b\}$, $a,b=0,1,\dots$.
Several explicit representations for Bethe vectors were found in \cite{BPRS-BV-12}. We present here one of them  using
the shorthand notation introduced in~\eqref{SH-prod},~\eqref{exam}
\begin{gather}\label{BV-explicit}
\mathbb{B}_{a,b}(\bar u;\bar v)=
\sum_{\substack{ {\bar u}\Rightarrow\{{\bar u}_{\so},{\bar u}_{\st}\}\\ {\bar v}\Rightarrow\{{\bar v}_{\so},{\bar v}_{\st}\}}}
\frac{{\mathsf{K}}_n(\bar v_{\so}|\bar u_{\so})}{ \lambda_2(\bar v_{\st})\lambda_2(\bar u)}
\frac{f(\bar v_{\st},\bar v_{\so})f(\bar u_{\so},\bar u_{\st})}
{f(\bar v,\bar u)}
  T_{13}(\bar u_{\so}) T_{12}(\bar u_{\st}) T_{23}(\bar v_{\st})|0\rangle.
\end{gather}
Here the sum is taken over partitions of the sets ${\bar u}\Rightarrow\{{\bar u}_{\so},{\bar u}_{\st}\}$ and ${\bar v}\Rightarrow\{{\bar v}_{\so},{\bar v}_{\st}\}$. The partitions are independent except the condition $\#{\bar u}_{\so}=\#{\bar v}_{\so}=n$, where $n=0,1,\dotsm\min(a,b)$.
The function ${\mathsf{K}}_n(\bar v_{\so}|\bar u_{\so})$ is a partition function of the six-vertex model with domain wall boundary conditions~\cite{Kor82}. It has the following explicit representation~\cite{Ize87}:
 \begin{gather*}%\label{K-def}
 {\mathsf{K}}_n(\bar v|\bar u)=\prod_{\ell<m}^n g(v_\ell,v_m)g(u_m,u_\ell)\cdot \frac{f(\bar v,\bar u)}{g(\bar v,\bar u)}
   \det \left.\left[ \frac{g^2(v_i,u_j)}{f(v_i,u_j)} \right]\right|_{i,j=1,\ldots,n}.
\end{gather*}

Partial Bethe vectors $\mathbb{B}_{a,b}^{(\ell)}(\bar u;\bar v)$ are given by the same formula \eqref{BV-explicit}, where one
should replace all $T_{ij}$ by $T_{ij}^{(\ell)}$,  all $\lambda_{2}$ by $\lambda_{2}^{(\ell)}$,  and $|0\rangle$  by $|0\rangle^{(\ell)}$,  $\ell=1,2$.
In order to express
the total Bethe vector in terms of matrix elements $T_{ij}^{(\ell)}$ acting on $|0\rangle^{(1)}\otimes |0\rangle^{(2)}$ one should substitute~\eqref{T-TT}
for every $T_{ij}$ and~\eqref{lr} for all~$\lambda_2$ into equation~\eqref{BV-explicit}. It is a highly nontrivial fact that the result of this substitution
can be reduced to a bilinear combination of partial Bethe vectors.

We denote dual Bethe vectors by $\mathbb{C}_{a,b}(\bar u;\bar v)$. They  are given by a formula similar to \eqref{BV-explicit}
\begin{gather*}%\label{dBV-explicit}
\mathbb{C}_{a,b}(\bar u;\bar v)=
\sum_{\substack{ {\bar u}\Rightarrow\{{\bar u}_{\so},{\bar u}_{\st}\}\\ {\bar v}\Rightarrow\{{\bar v}_{\so},{\bar v}_{\st}\}}}
\frac{{\mathsf{K}}_n(\bar v_{\so}|\bar u_{\so})}{ \lambda_2(\bar v_{\st})\lambda_2(\bar u)}
\frac{f(\bar v_{\st},\bar v_{\so})f(\bar u_{\so},\bar u_{\st})}
{f(\bar v,\bar u)}
\langle0| T_{32}(\bar v_{\st}) T_{21}(\bar u_{\st})T_{31}(\bar u_{\so}).
\end{gather*}
Here the notation is the same as in \eqref{BV-explicit}. In order to obtain partial dual Bethe vectors one should make the replacements
already mentioned, namely $T_{ij}\to T_{ij}^{(\ell)}$, $\lambda_{2}\to \lambda_{2}^{(\ell)}$, and $\langle0|\to \langle0|^{(\ell)}$.

\subsection{Morphisms of the algebra and relations between Bethe vectors}

The $R$-matrix \eqref{R-mat} is invariant under simultaneous transpositions in both auxiliary spaces. This fact implies the existence
of two symmetries in the algebra of the monodromy matrix elements. The mapping
\begin{gather}\label{psi-m}
\psi\colon \  T_{ij}(u) \to   T_{ji}(u)
\end{gather}
def\/ines an antimorphism of the algebra \eqref{RTT-al}, while the mapping
\begin{gather}\label{phi-m}
\varphi\colon \  T_{ij}(u)  \to  T_{4-j,4-i}(-u)
\end{gather}
def\/ines an isomorphism of the algebra~\eqref{RTT-al}.

Assuming that
\begin{gather*}%\label{vac-ma}
\psi(|0\rangle)=\langle 0|,\qquad \psi(\langle 0|)=|0\rangle,\qquad \varphi(|0\rangle)=|0\rangle,\qquad \varphi(\langle 0|)=\langle 0|
\end{gather*}
we obtain from the mappings \eqref{psi-m} and \eqref{phi-m} the following relations between dif\/ferent Bethe vectors \cite{BPRS-BV-12}
\begin{gather}\label{psi-BV}
\psi(\mathbb{B}_{a,b}({\bar u};{\bar v}))= \mathbb{C}_{a,b}({\bar u};{\bar v}),\qquad \psi(\mathbb{C}_{a,b}({\bar u};{\bar v}))= \mathbb{B}_{a,b}({\bar u};{\bar v}),
\end{gather}
and
\begin{gather*}%\label{phi-BV}
\varphi(\mathbb{B}_{a,b}({\bar u};{\bar v}))= \mathbb{B}_{b,a}(-{\bar v};-{\bar u}),\qquad
\varphi(\mathbb{C}_{a,b}({\bar u};{\bar v}))= \mathbb{C}_{b,a}(-{\bar v};-{\bar u}).
\end{gather*}

Moreover the mappings \eqref{psi-m} and \eqref{phi-m} intertwine the coproduct \eqref{uni-copro} and inverse coproduct
\begin{gather*}%\label{inv-un-co}
\Delta' T_{ij}(u)=\sum_{k=1}^{3} T_{ik}(u)\otimes T_{kj}(u).
\end{gather*}
Namely
\begin{gather}\label{int-psi-phi}
\Delta\circ \psi =(\psi\otimes \psi)\circ \Delta',\qquad  \Delta\circ \varphi =(\varphi\otimes \varphi)\circ \Delta'.
\end{gather}
For example, the f\/irst intertwining relation follows from the chain of equalities
\begin{gather*}
\Delta\circ \psi (T_{ij}(u))=\Delta(T_{ji}(u))=\sum_{k=1}^3 T_{ki}(u)\otimes T_{jk}(u) \\
\hphantom{\Delta\circ \psi (T_{ij}(u))}{}
= \sum_{k=1}^3 \psi(T_{ik}(u))\otimes \psi(T_{kj}(u))=
(\psi\otimes \psi)\circ \Delta' (T_{ij}(u)).
\end{gather*}
The second property in~\eqref{int-psi-phi} can be proved  in the same way.  Thus, we see that the action of the
mappings~$\psi$ and~$\varphi$ exchanges the components in the tensor product.

\subsection{Main statement}\label{SS-MS}

\begin{thm}\label{main-theor}
The Bethe vectors of the  total monodromy matrix $T(u)$ can be presented as a~bilinear combination of partial Bethe vectors
as follows:
\begin{gather}\label{BV-BV}
\mathbb{B}_{a,b}({\bar u};{\bar v})=\sum r_{1}^{(2)}({\bar u}_{\so}) r_{3}^{(1)}({\bar v}_{\st})\frac{f({\bar u}_{\st},{\bar u}_{\so})f({\bar v}_{\st},{\bar v}_{\so})}{f({\bar v}_{\st},{\bar u}_{\so})}
\mathbb{B}_{a_1,b_1}^{(1)}({\bar u}_{\so};{\bar v}_{\so}) \mathbb{B}_{a_2,b_2}^{(2)}({\bar u}_{\st};{\bar v}_{\st}).
\end{gather}
Here the sums are taken over  all the partitions ${\bar u}\Rightarrow\{{\bar u}_{\so},{\bar u}_{\st}\}$ and ${\bar v}\Rightarrow\{{\bar v}_{\so},{\bar v}_{\st}\}$. The integers
$a_\ell$ and $b_\ell$ $(\ell=1,2)$ are the cardinalities of the corresponding subsets. Hereby, $a_1+a_2=a$ and $b_1+b_2=b$.
\end{thm}

\begin{proof}[Sketch of proof]
 We use an induction over $a$. First, for arbitrary $a, b\ge0$ we def\/ine a vec\-tor~$\mathcal{B}_{a,b}({\bar u},{\bar v})$ as follows:
\begin{gather}\label{New-BV}
\mathcal{B}_{a,b}({\bar u};{\bar v})=\sum r_{1}^{(2)}({\bar u}_{\so}) r_{3}^{(1)}({\bar v}_{\st})\frac{f({\bar u}_{\st},{\bar u}_{\so})f({\bar v}_{\st},{\bar v}_{\so})}{f({\bar v}_{\st},{\bar u}_{\so})}
\mathbb{B}_{a_1,b_1}^{(1)}({\bar u}_{\so};{\bar v}_{\so}) \mathbb{B}_{a_2,b_2}^{(2)}({\bar u}_{\st};{\bar v}_{\st}).
\end{gather}
When $a=0$ and  $b$ is arbitrary, we obtain
\begin{gather}\label{BV-BVa0}
\mathcal{B}_{0,b}(\varnothing;{\bar v})=\sum  r_{3}^{(1)}({\bar v}_{\st})f({\bar v}_{\st},{\bar v}_{\so})
\mathbb{B}_{0,b_1}^{(1)}(\varnothing;{\bar v}_{\so}) \mathbb{B}_{0,b_2}^{(2)}(\varnothing;{\bar v}_{\st}).
\end{gather}
This is a known formula for the Bethe vector in the composite model in the ${\rm GL}(2)$ case \cite{IzK84}.
Thus, we conclude that $\mathcal{B}_{0,b}(\varnothing;{\bar v})=\mathbb{B}_{0,b}(\varnothing;{\bar v})$.

Then we consider the action of the operators $T_{13}(z)$ and $T_{12}(z)$ onto this vector. The goal  is to prove that
\begin{gather}\label{act-T13}
\frac{T_{13}(z)}{\lambda_2(z)}\mathcal{B}_{a-1,b-1}({\bar u};{\bar v}) =\mathcal{B}_{a,b}(\{{\bar u},z\};\{{\bar v},z\}),
\end{gather}
and
\begin{gather}\label{act-T12}
\frac{T_{12}(z)}{\lambda_2(z)}\mathcal{B}_{a-1,b}({\bar u};{\bar v}) =f({\bar v},z)\mathcal{B}_{a,b}(\{{\bar u},z\};{\bar v})
+\sum g(z,v_0)f({\bar v}_0,v_0) \mathcal{B}_{a,b}(\{{\bar u},z\};\{{\bar v}_0,z\}),\!\!\!
\end{gather}
where the sum is taken over partitions ${\bar v}\Rightarrow\{v_0,{\bar v}_0\}$ with $\# v_0=1$  (recall that due to our
convention ${\bar v}_0={\bar v}\setminus v_0$).

If these actions are proved, then we obtain a recursion
\begin{gather}\label{recursion}
\mathcal{B}_{a,b}(\{{\bar u},z\};{\bar v})=\frac{T_{12}(z)\mathcal{B}_{a-1,b}({\bar u};{\bar v})}{\lambda_2(z)f({\bar v},z)}
+\sum g(z,v_0)f({\bar v}_0,v_0) \frac{T_{13}(z)\mathcal{B}_{a-1,b-1}({\bar u};{\bar v}_0)}{\lambda_2(z)f({\bar v},z)}.
\end{gather}
Since this recursion coincides with the one for the Bethe vectors  (see, e.g.,~\cite{BPRS-BV-12}),
and using the equality proven above for $a=0$, we conclude that  $\mathcal{B}_{a,b}$ is
a Bethe vector~$\mathbb{B}_{a,b}$.
 Thus, the proof of Theorem~\ref{main-theor} reduces to proving equations~\eqref{act-T13} and~\eqref{act-T12}: it is done in Sections~\ref{T13} and~\ref{T12}.
\end{proof}

\begin{cor}\label{main-cor}
Dual Bethe vectors of the total monodromy matrix $T(u)$ can be presented as a~bilinear combination of dual partial Bethe vectors
as follows:
\begin{gather}\label{BV-BV-du}
\mathbb{C}_{a,b}({\bar u};{\bar v})=\sum r_{1}^{(1)}({\bar u}_{\st}) r_{3}^{(2)}({\bar v}_{\so})\frac{f({\bar u}_{\so},{\bar u}_{\st})f({\bar v}_{\so},{\bar v}_{\st})}{f({\bar v}_{\so},{\bar u}_{\st})}
\mathbb{C}_{a_1,b_1}^{(1)}({\bar u}_{\so};{\bar v}_{\so}) \mathbb{C}_{a_2,b_2}^{(2)}({\bar u}_{\st};{\bar v}_{\st}).
\end{gather}
Here the sums are taken over partitions ${\bar u}\Rightarrow\{{\bar u}_{\so},{\bar u}_{\st}\}$ and ${\bar v}\Rightarrow\{{\bar v}_{\so},{\bar v}_{\st}\}$. The integers
$a_\ell$ and $b_\ell$ $(\ell=1,2)$ are the cardinalities of the corresponding subsets. Hereby, $a_1+a_2=a$ and $b_1+b_2=b$.
\end{cor}

\begin{proof} Using \eqref{psi-BV} we  act  with the antimorphism $\psi$ on \eqref{BV-BV}. Since the action of
$\psi$ exchanges the components of the tensor product, we  should replace in the r.h.s.\ the vector $\mathbb{B}^{(1)}$ by $\mathbb{C}^{(2)}$
and the vector $\mathbb{B}^{(2)}$ by $\mathbb{C}^{(1)}$. The subtlety is that we also should make the replacement of the functions $r_k^{(\ell)}\colon r_k^{(1)}\leftrightarrow r_k^{(2)}$. This can be easily understood if we remember the origin of these functions in~\eqref{BV-BV}.
Due to \eqref{BV-explicit} total Bethe vectors are polynomials in operators $T_{ij}$ with $i<j$ acting on the vacuum vector
\begin{gather*}%\label{BV-polyn}
\mathbb{B}_{a,b}({\bar u};{\bar v})=P(T_{ij})|0\rangle,
\end{gather*}
which implies
\begin{gather*}%\label{BV-polyn-1}
\mathbb{B}_{a,b}({\bar u};{\bar v})=P\left(\sum_{k=1}^3T^{(2)}_{ik}T^{(1)}_{kj}\right)|0\rangle.
\end{gather*}
Therefore, in spite of the total Bethe vector depends only on $T_{ij}$ with $i<j$, it depends on the partial $T^{(\ell)}_{ij}$ with $i=j$
as well. It is the action of these operators on the vacuum vector that produces the functions $r_i^{(\ell)}$. But since the action of
$\psi$ exchanges the components of the tensor product, we obtain that $T^{(1)}_{ii}\leftrightarrow  T^{(2)}_{ii}$, and hence,
$r_i^{(1)}\leftrightarrow r_i^{(2)}$. Thus, we obtain
\begin{gather*}%\label{BV-BV-predu}
\mathbb{C}_{a,b}({\bar u};{\bar v})=\sum r_{1}^{(1)}({\bar u}_{\so}) r_{3}^{(2)}({\bar v}_{\st})\frac{f({\bar u}_{\st},{\bar u}_{\so})f({\bar v}_{\st},{\bar v}_{\so})}{f({\bar v}_{\st},{\bar u}_{\so})}
\mathbb{C}_{a_2,b_2}^{(1)}({\bar u}_{\st};{\bar v}_{\st}) \mathbb{C}_{a_1,b_1}^{(2)}({\bar u}_{\so};{\bar v}_{\so}).
\end{gather*}
It remains to relabel the subsets ${\bar u}_{\so} \leftrightarrow {\bar u}_{\st}$, ${\bar v}_{\so} \leftrightarrow {\bar v}_{\st}$ and the
subscripts of the partial dual Bethe vectors $a_1 \leftrightarrow a_2$, $b_1 \leftrightarrow b_2$, and we arrive at~\eqref{BV-BV-du}.
\end{proof}

\section[Action of $T_{13}(z)$]{Action of $\boldsymbol{T_{13}(z)}$}\label{T13}

In order to study the action of the operators $T_{12}(u)$ and $T_{13}(u)$ onto partial Bethe vectors we
use \eqref{T-TT} and present these operators in the form
\begin{gather}
T_{12}(u) =T_{11}^{(2)}(u)T_{12}^{(1)}(u)+T_{12}^{(2)}(u)T_{22}^{(1)}(u)+T_{13}^{(2)}(u)T_{32}^{(1)}(u),\nonumber\\
T_{13}(u) =T_{11}^{(2)}(u)T_{13}^{(1)}(u)+T_{12}^{(2)}(u)T_{23}^{(1)}(u)+T_{13}^{(2)}(u)T_{33}^{(1)}(u).\label{T12T13}
\end{gather}

Let $\{z,{\bar u}\}={\bar\eta}$ and $\{z,{\bar v}\}={\bar\xi}$.  Then due to \eqref{act-T13} we should prove that
\begin{gather}\label{act-T13z}
\frac{T_{13}(z)}{\lambda_2(z)}\mathcal{B}_{a-1,b-1}({\bar u};{\bar v}) =\mathcal{B}_{a,b}({\bar\eta};{\bar\xi}).
\end{gather}
Due to \eqref{New-BV} a vector $\mathcal{B}_{a,b}({\bar\eta};{\bar\xi})$ has the form
\begin{gather}\label{NBV}
\mathcal{B}_{a,b}({\bar\eta};{\bar\xi})=\sum r_{1}^{(2)}({\bar\eta}_{\so}) r_{3}^{(1)}({\bar\xi}_{\st})\frac{f({\bar\eta}_{\st},{\bar\eta}_{\so})f({\bar\xi}_{\st},{\bar\xi}_{\so})}{f({\bar\xi}_{\st},{\bar\eta}_{\so})}
\mathbb{B}^{(1)}({\bar\eta}_{\so};{\bar\xi}_{\so}) \mathbb{B}^{(2)}({\bar\eta}_{\st};{\bar\xi}_{\st}).
\end{gather}

\begin{Rem} We have omitted the subscripts of the partial Bethe vectors in the r.h.s.\ of~\eqref{NBV},
because in this case they do not give any
additional information. Indeed, the subscripts of partial Bethe vectors
are equal to the cardinalities of the subsets of Bethe parameters. Since  in~\eqref{NBV}, the sum over partitions
is taken over all possible subsets, it is clear that the corresponding cardinalities run through all possible
values from~$0$ to~$a$ for the subsets of~${\bar u}$ and  from~$0$ to~$b$ for the subsets of~${\bar v}$. Therefore below
we shall omit the subscripts of the partial Bethe vectors in the sums over partitions.
\end{Rem}

Consider how the parameter $z$ may enter the subsets of ${\bar\eta}$ and ${\bar\xi}$. Obviously, there are three cases in the r.h.s.:
\begin{alignat*}{6}
& {\rm (i)}\quad &&{\bar\eta}_{\so}=\{z,{\bar u}_{\so}\}, \qquad && {\bar\xi}_{\so}=\{z,{\bar v}_{\so}\}, \qquad && {\bar\eta}_{\st}={\bar u}_{\st},   \qquad&& {\bar\xi}_{\st}={\bar v}_{\st},& \nonumber\\
& {\rm (ii)}\quad && {\bar\eta}_{\so}={\bar u}_{\so}, \qquad && {\bar\xi}_{\so}={\bar v}_{\so},\qquad && {\bar\eta}_{\st}=\{z,{\bar u}_{\st}\},  \qquad && {\bar\xi}_{\st}=\{z,{\bar v}_{\st}\},&\nonumber\\
& {\rm (iii)}\quad && {\bar\eta}_{\so}={\bar u}_{\so},  \qquad && {\bar\xi}_{\so}=\{z,{\bar v}_{\so}\},  \qquad && {\bar\eta}_{\st}=\{z,{\bar u}_{\st}\}, \qquad &&  {\bar\xi}_{\st}={\bar v}_{\st}.&%\label{cases}
\end{alignat*}
The case $z\in{\bar\eta}_{\so}$ and $z\in{\bar\xi}_{\st}$ gives vanishing contribution,  because  the product
$f({\bar\xi}_{\st},{\bar\eta}_{\so})^{-1}$  contains the factor $f(z,z)^{-1}=0$. Thus, we obtain
\begin{gather*}%\label{NBV-3terms}
\mathcal{B}_{a,b}({\bar\eta};{\bar\xi})=A_1+A_2+A_3,
\end{gather*}
where
\begin{gather}
A_1 = \sum r_{1}^{(2)}(z)r_{1}^{(2)}({\bar u}_{\so}) r_{3}^{(1)}({\bar v}_{\st})\frac{f({\bar u}_{\st},z)f({\bar u}_{\st},{\bar u}_{\so})f({\bar v}_{\st},{\bar v}_{\so})}{f({\bar v}_{\st},{\bar u}_{\so})} \nonumber\\
\hphantom{A_1 =}{}\times
 \mathbb{B}^{(1)}(\{{\bar u}_{\so},z\};\{{\bar v}_{\so},z\}) \mathbb{B}^{(2)}({\bar u}_{\st};{\bar v}_{\st}),
\label{A1}
\\
A_2 = \sum r_{3}^{(1)}(z) r_{1}^{(2)}({\bar u}_{\so}) r_{3}^{(1)}({\bar v}_{\st})\frac{f({\bar u}_{\st},{\bar u}_{\so})f({\bar v}_{\st},{\bar v}_{\so})f(z,{\bar v}_{\so})}
{f({\bar v}_{\st},{\bar u}_{\so})} \nonumber\\
\hphantom{A_2 =}{}\times
 \mathbb{B}^{(1)}({\bar u}_{\so};{\bar v}_{\so}) \mathbb{B}^{(2)}(\{{\bar u}_{\st},z\};\{{\bar v}_{\st},z\}),
\label{A2}
\\
A_3 = \sum r_{1}^{(2)}({\bar u}_{\so}) r_{3}^{(1)}({\bar v}_{\st})\frac{f(z,{\bar u}_{\so})f({\bar v}_{\st},z)
f({\bar u}_{\st},{\bar u}_{\so})f({\bar v}_{\st},{\bar v}_{\so})}
{f({\bar v}_{\st},{\bar u}_{\so})} \nonumber\\
\hphantom{A_3 =}{}\times
 \mathbb{B}^{(1)}({\bar u}_{\so};\{{\bar v}_{\so},z\}) \mathbb{B}^{(2)}(\{{\bar u}_{\st},z\};{\bar v}_{\st}).
\label{A3}
\end{gather}

Consider now the action of the operator $T_{13}(z)$ onto the vector $\mathcal{B}_{a-1,b-1}({\bar u};{\bar v})$ in the l.h.s.\ of~\eqref{act-T13z}. Due to
\eqref{T12T13}  we have
\begin{gather*}
\frac{T_{13}(z)}{\lambda_2(z)}\mathcal{B}_{a-1,b-1}({\bar u};{\bar v}) =
\left(\frac{T_{11}^{(2)}(z)T_{13}^{(1)}(z)}{\lambda_2^{(2)}(z)\lambda_2^{(1)}(z)}
+\frac{T_{12}^{(2)}(z)T_{23}^{(1)}(z)}{\lambda_2^{(2)}(z)\lambda_2^{(1)}(z)}
+\frac{T_{13}^{(2)}(z)T_{33}^{(1)}(z)}{\lambda_2^{(2)}(z)\lambda_2^{(1)}(z)}\right)
\nonumber\\
\hphantom{\frac{T_{13}(z)}{\lambda_2(z)}\mathcal{B}_{a-1,b-1}({\bar u};{\bar v}) = }{} \times \mathcal{B}_{a-1,b-1}({\bar u};{\bar v}).%\label{act-T13NBV}
\end{gather*}
Substituting here \eqref{New-BV} for $\mathcal{B}_{a-1,b-1}({\bar u};{\bar v})$ we f\/ind
\begin{gather*}%\label{act-T13NBV-3t}
\frac{T_{13}(z)}{\lambda_2(z)}\mathcal{B}_{a-1,b-1}({\bar u};{\bar v})=C_1+C_2+C_3,
\end{gather*}
where
\begin{gather*}
 C_k=\sum r_{1}^{(2)}({\bar u}_{\so}) r_{3}^{(1)}({\bar v}_{\st})\frac{f({\bar u}_{\st},{\bar u}_{\so})f({\bar v}_{\st},{\bar v}_{\so})}{f({\bar v}_{\st},{\bar u}_{\so})}
\frac{T_{k3}^{(1)}(z)}{\lambda_2^{(1)}(z)}\mathbb{B}^{(1)}({\bar u}_{\so};{\bar v}_{\so})
\frac{T_{1k}^{(2)}(z)}{\lambda_2^{(2)}(z)}\mathbb{B}^{(2)}({\bar u}_{\st};{\bar v}_{\st}),
\nonumber\\
  k=1,2,3.
%\label{C123}
\end{gather*}

Using formulas for the action of the monodromy matrix elements onto Bethe vectors (see Appendix~\ref{Action})
we can f\/ind the terms $C_k$ explicitly.

 Due to  \eqref{A-13} and \eqref{A-11} we have
\begin{gather}
C_1
=\sum_{\substack{{\bar u}\Rightarrow\{{\bar u}_{\so},{\bar u}_{\st}\}\\{\bar v}\Rightarrow\{{\bar v}_{\so},{\bar v}_{\st}\}} }
r_{1}^{(2)}({\bar u}_{\so}) r_{3}^{(1)}({\bar v}_{\st})\frac{f({\bar u}_{\st},{\bar u}_{\so})f({\bar v}_{\st},{\bar v}_{\so})}{f({\bar v}_{\st},{\bar u}_{\so})}
\mathbb{B}^{(1)}(\{{\bar u}_{\so},z\};\{{\bar v}_{\so},z\})\nonumber\\
\hphantom{C_1=}{}
\times\Biggl(r_1^{(2)}(z)f({\bar u}_{\st},z)\mathbb{B}^{(2)}({\bar u}_{\st};{\bar v}_{\st}) \nonumber\\
\hphantom{C_1=}{}
 + f({\bar v}_{\st},z)\sum_{{\bar u}_{\st}\Rightarrow\{u_{\rm i},{\bar u}_{\rm ii}\}} {  r_1^{(2)}(u_{\rm i})g(z,u_{\rm i})\frac{f({\bar u}_{\rm ii},u_{\rm i})}
{f({\bar v}_{\st}, u_{\rm i})}\mathbb{B}^{(2)}(\{{\bar u}_{\rm ii},z\};{\bar v}_{\st})}\nonumber\\
\hphantom{C_1=}{}
 +\sum_{\substack{{\bar u}_{\st}\Rightarrow\{u_{\rm i},{\bar u}_{\rm ii}\}\\{\bar v}_{\st}\Rightarrow\{v_{\rm i},{\bar v}_{\rm ii}\}}} r_1^{(2)}(u_{\rm i})
g(z,v_{\rm i})g(v_{\rm i},u_{\rm i})\frac{f({\bar u}_{\rm ii},u_{\rm i})f({\bar v}_{\rm ii},v_{\rm i})}{f({\bar v}_{\st}, u_{\rm i})}
\mathbb{B}^{(2)}(\{{\bar u}_{\rm ii},z\};\{{\bar v}_{\rm ii},z\})\Biggr).\label{C1}
\end{gather}
Here the original sets of the Bethe parameters are divided into subsets ${\bar u}\Rightarrow\{{\bar u}_{\so},{\bar u}_{\st}\}$ and ${\bar v}\Rightarrow\{{\bar v}_{\so},{\bar v}_{\st}\}$. Then in some terms the subsets ${\bar u}_{\st}$ and ${\bar v}_{\st}$ are divided into
additional subsets ${\bar u}_{\st}\Rightarrow\{u_{\rm i},{\bar u}_{\rm ii}\}$ and ${\bar v}_{\st}\Rightarrow\{v_{\rm i},{\bar v}_{\rm ii}\}$
with $\#u_{\rm i}=\#v_{\rm i}=1$ (for this reason we do not write bar for~$u_{\rm i}$ and~$v_{\rm i}$). The sum is taken over all
partitions described above.

Similarly, due to \eqref{A-13}, \eqref{A-12}, \eqref{A-23}, and \eqref{A-33} we obtain
\begin{gather}
C_2
=\sum_{\substack{{\bar u}\Rightarrow\{{\bar u}_{\so},{\bar u}_{\st}\}\\{\bar v}\Rightarrow\{{\bar v}_{\so},{\bar v}_{\st}\}} }
r_{1}^{(2)}({\bar u}_{\so}) r_{3}^{(1)}({\bar v}_{\st})\frac{f({\bar u}_{\st},{\bar u}_{\so})f({\bar v}_{\st},{\bar v}_{\so})}{f({\bar v}_{\st},{\bar u}_{\so})}\nonumber\\
\hphantom{C_2=}{}
\times \Biggl(f(z,{\bar u}_{\so})\mathbb{B}^{(1)}({\bar u}_{\so};\{{\bar v}_{\so},z\})
+{\sum_{{\bar u}_{\so}\Rightarrow\{u_{\rm i},{\bar u}_{\rm ii}\}}g(u_{\rm i},z)f(u_{\rm i},{\bar u}_{\rm ii})
\mathbb{B}^{(1)}(\{{\bar u}_{\rm ii},z\};\{{\bar v}_{\so},z\})}\Biggr)\nonumber\\
\hphantom{C_2=}{}
\times\Biggl(f({\bar v}_{\st},z)\mathbb{B}^{(2)}(\{{\bar u}_{\st},z\};{\bar v}_{\st})
+{\sum_{{\bar v}_{\st}\Rightarrow\{v_{\rm i},{\bar v}_{\rm ii}\}}g(z,v_{\rm i})f({\bar v}_{\rm ii},v_{\rm i})
\mathbb{B}^{(2)}(\{{\bar u}_{\st},z\};\{{\bar v}_{\rm ii},z\})}\Biggr),\label{C2}
\end{gather}
and
\begin{gather}
C_3
=\sum_{\substack{{\bar u}\Rightarrow\{{\bar u}_{\so},{\bar u}_{\st}\}\\{\bar v}\Rightarrow\{{\bar v}_{\so},{\bar v}_{\st}\}} }
r_{1}^{(2)}({\bar u}_{\so}) r_{3}^{(1)}({\bar v}_{\st})\frac{f({\bar u}_{\st},{\bar u}_{\so})f({\bar v}_{\st},{\bar v}_{\so})}{f({\bar v}_{\st},{\bar u}_{\so})}
\Biggl(r_3^{(1)}(z)f(z,{\bar v}_{\so})\mathbb{B}^{(1)}({\bar u}_{\so};{\bar v}_{\so}) \nonumber\\
\hphantom{C_3=}{} +
{f(z,{\bar u}_{\so})\sum_{{\bar v}_{\so}\Rightarrow\{v_{\rm i},{\bar v}_{\rm ii}\}}r_3^{(1)}(v_{\rm i})g(v_{\rm i},z)\frac{f(v_{\rm i},{\bar v}_{\rm ii})}{f(v_{\rm i},{\bar u}_{\so})}\mathbb{B}^{(1)}({\bar u}_{\so};\{{\bar v}_{\rm ii},z\})}\nonumber\\
\hphantom{C_3=}{}
 +\sum_{\substack{{\bar u}_{\so}\Rightarrow\{u_{\rm i},{\bar u}_{\rm ii}\}\\{\bar v}_{\so}\Rightarrow\{v_{\rm i},{\bar v}_{\rm ii}\}}}r_3^{(1)}(v_{\rm i})
g(u_{\rm i},z)g(v_{\rm i},u_{\rm i})\frac{f(u_{\rm i},{\bar u}_{\rm ii})f(v_{\rm i},{\bar v}_{\rm ii})}{f(v_{\rm i},{\bar u}_{\so})}\mathbb{B}^{(1)}(\{{\bar u}_{\rm ii},z\};\{{\bar v}_{\rm ii},z\})\Biggr)\nonumber\\
\hphantom{C_3=}{}
\times
\mathbb{B}^{(2)}(\{{\bar u}_{\st},z\};\{{\bar v}_{\st},z\}).\label{C3}
\end{gather}

Looking at~\eqref{C1} we see that the sum over partitions involving the product of Bethe vectors $\mathbb{B}^{(1)}(\{{\bar u}_{\so},z\};\{{\bar v}_{\so},z\})\mathbb{B}^{(2)}({\bar u}_{\st};{\bar v}_{\st})$ coincides with the term~$A_1$~\eqref{A1}.
Similarly, the sum in \eqref{C2} involving the product
$\mathbb{B}^{(1)}({\bar u}_{\so};\{{\bar v}_{\so},z\})\mathbb{B}^{(2)}(\{{\bar u}_{\st},z\};{\bar v}_{\st})$ coincides with~$A_3$~\eqref{A3},
while the sum in~\eqref{C3} involving the product
$\mathbb{B}^{(1)}({\bar u}_{\so};{\bar v}_{\so})\mathbb{B}^{(2)}(\{{\bar u}_{\st},z\};\{{\bar v}_{\st},z\})$ coincides with $A_2$ \eqref{A2}.
Thus, equation~\eqref{act-T13} will be proved if we show that contributions from other  terms in  \eqref{C1}--\eqref{C3} va\-nish.
To observe these cancellations we combine together the terms where the Bethe vectors~$\mathbb{B}^{(1)}$ and~$\mathbb{B}^{(2)}$ depend on the
parameter~$z$ in the same manner.

There are  two terms  containing  the product of the following type:
\begin{gather*}%\label{1-type}
\mathbb{B}^{(1)}(\{{\bar u}',z\};\{{\bar v}',z\})\mathbb{B}^{(2)}(\{{\bar u}'',z\};\{{\bar v}''\}).
\end{gather*}
Here  ${\bar u}'$, ${\bar u}''$ and ${\bar v}'$, ${\bar v}''$ are arbitrary subsets of the sets ${\bar u}$ and ${\bar v}$ respectively. The f\/irst term
of this type comes from \eqref{C1}
\begin{gather}
C_{12}=\sum_{\substack{{\bar u}\Rightarrow\{{\bar u}_{\so},u_{\rm i},{\bar u}_{\rm ii}\}\\{\bar v}\Rightarrow\{{\bar v}_{\so},{\bar v}_{\st}\}} }
r_1^{(2)}(u_{\rm i})r_{1}^{(2)}({\bar u}_{\so}) r_{3}^{(1)}({\bar v}_{\st})
\frac{f({\bar u}_{\rm ii},{\bar u}_{\so})f(u_{\rm i},{\bar u}_{\so})f({\bar v}_{\st},{\bar v}_{\so})f({\bar v}_{\st},z)f({\bar u}_{\rm ii},u_{\rm i})}
{f({\bar v}_{\st},{\bar u}_{\so})f({\bar v}_{\st}, u_{\rm i})}\nonumber\\
\hphantom{C_{12}=}{}
\times {g(z,u_{\rm i})} \mathbb{B}^{(1)}(\{{\bar u}_{\so},z\};\{{\bar v}_{\so},z\})\mathbb{B}^{(2)}(\{{\bar u}_{\rm ii},z\};{\bar v}_{\st}).\label{1-type-1}
\end{gather}
The second term is contained in \eqref{C2}:
\begin{gather}
C_{22}=\sum_{\substack{{\bar u}\Rightarrow\{u_{\rm i},{\bar u}_{\rm ii},{\bar u}_{\st}\}\\{\bar v}\Rightarrow\{{\bar v}_{\so},{\bar v}_{\st}\}} }
r_{1}^{(2)}(u_{\rm i})r_{1}^{(2)}({\bar u}_{\so}) r_{3}^{(1)}({\bar v}_{\st}) \frac{f({\bar u}_{\st},u_{\rm i})f({\bar u}_{\st},{\bar u}_{\rm ii})
f(u_{\rm i},{\bar u}_{\rm ii})f({\bar v}_{\st},{\bar v}_{\so})f({\bar v}_{\st},z)}
{f({\bar v}_{\st},u_{\rm i})f({\bar v}_{\st},{\bar u}_{\rm ii})}\nonumber\\
\hphantom{C_{22}=}{}
\times g(u_{\rm i},z) \mathbb{B}^{(1)}(\{{\bar u}_{\rm ii},z\};\{{\bar v}_{\so},z\})\mathbb{B}^{(2)}(\{{\bar u}_{\st},z\};{\bar v}_{\st}).\label{1-type-2}
\end{gather}
We claim that $C_{12}+C_{22}=0$. For this purpose we relabel the subsets in  equation~\eqref{1-type-2}, so that the vectors in~\eqref{1-type-1} and~\eqref{1-type-2} to have the same arguments. Observe that such the replacement is nothing but the change of summation variables.
Clearly, we should make the following relabeling:
f\/irst ${\bar u}_{\rm ii}\to {\bar u}_{\so}$ and then ${\bar u}_{\st}\to{\bar u}_{\rm ii}$
\begin{gather*}
\mathbb{B}^{(1)}(\{{\bar u}_{\rm ii},z\};\{{\bar v}_{\so},z\})\mathbb{B}^{(2)}(\{{\bar u}_{\st},z\};{\bar v}_{\st})\xrightarrow{{\bar u}_{\rm ii}\to {\bar u}_{\so}}
\mathbb{B}^{(1)}(\{{\bar u}_{\so},z\};\{{\bar v}_{\so},z\})\mathbb{B}^{(2)}(\{{\bar u}_{\st},z\};{\bar v}_{\st}),\nonumber\\
\mathbb{B}^{(1)}(\{{\bar u}_{\so},z\};\{{\bar v}_{\so},z\})\mathbb{B}^{(2)}(\{{\bar u}_{\st},z\};{\bar v}_{\st})\xrightarrow{{\bar u}_{\st}\to{\bar u}_{\rm ii}}
\mathbb{B}^{(1)}(\{{\bar u}_{\so},z\};\{{\bar v}_{\so},z\})\mathbb{B}^{(2)}(\{{\bar u}_{\rm ii},z\};{\bar v}_{\st}).%\label{relabel}
\end{gather*}
Then we obtain
\begin{gather*}%\label{1-type-22}
C_{22}=\sum_{\substack{{\bar u}\Rightarrow\{u_{\rm i},{\bar u}_{\so},{\bar u}_{\rm ii}\}\\{\bar v}\Rightarrow\{{\bar v}_{\so},{\bar v}_{\st}\}} }
r_{1}^{(2)}(u_{\rm i})r_{1}^{(2)}({\bar u}_{\so}) r_{3}^{(1)}({\bar v}_{\st}) \frac{f({\bar u}_{\rm ii},u_{\rm i})f({\bar u}_{\rm ii},{\bar u}_{\so})
f(u_{\rm i},{\bar u}_{\so})f({\bar v}_{\st},{\bar v}_{\so})f({\bar v}_{\st},z)}
{f({\bar v}_{\st},u_{\rm i})f({\bar v}_{\st},{\bar u}_{\so})}\\
\hphantom{C_{22}=}{}
\times g(u_{\rm i},z) \mathbb{B}^{(1)}(\{{\bar u}_{\so},z\};\{{\bar v}_{\so},z\})\mathbb{B}^{(2)}(\{{\bar u}_{\rm ii},z\};{\bar v}_{\st}).
\end{gather*}
We see that  $C_{12}+C_{22}=0$ due to the trivial identity
\begin{gather}\label{1-t-id}
g(z,u_{\rm i})+g(u_{\rm i},z)=0 .
\end{gather}

The analysis of other contributions can be done in the similar manner. There are two terms  containing the sums over
partitions involving  the products of the Bethe vectors of the form
\begin{gather*}%\label{2-type}
\mathbb{B}^{(1)}(\{{\bar u}'\};\{{\bar v}',z\})\mathbb{B}^{(2)}(\{{\bar u}'',z\};\{{\bar v}'',z\}).
\end{gather*}
It is easy to check that after appropriate relabeling of the subsets they also cancel
each other due to identity~\eqref{1-t-id}.

Finally, all three terms  \eqref{C1}--\eqref{C3} contain the sums over
partitions involving  the products of the Bethe vectors of the form
\begin{gather*}%\label{3-type}
\mathbb{B}^{(1)}(\{{\bar u}',z\};\{{\bar v}',z\})\mathbb{B}^{(2)}(\{{\bar u}'',z\};\{{\bar v}'',z\}).
\end{gather*}
Mutual cancellation of these terms is a  bit more subtle.
These contributions are
\begin{gather*}%\label{3-type-1}
C_{13}=\sum_{\substack{{\bar u}\Rightarrow\{{\bar u}_{\so},u_{\rm i},{\bar u}_{\rm ii}\}\\{\bar v}\Rightarrow\{{\bar v}_{\so},v_{\rm i},{\bar v}_{\rm ii}\}} }
r_1^{(2)}(u_{\rm i})r_{1}^{(2)}({\bar u}_{\so}) r_{3}^{(1)}({\bar v}_{\rm ii})r_{3}^{(1)}(v_{\rm i})
\frac{f(u_{\rm i},{\bar u}_{\so})f({\bar u}_{\rm ii},{\bar u}_{\so})f({\bar u}_{\rm ii},u_{\rm i})}
{f(v_{\rm i},{\bar u}_{\so})f({\bar v}_{\rm ii},{\bar u}_{\so})f({\bar v}_{\rm ii},{\bar u}_{\so})f(v_{\rm i}, u_{\rm i})}
 \\
\hphantom{C_{13}=}{}
\times f({\bar v}_{\rm ii},v_{\rm i})f(v_{\rm i},{\bar v}_{\so})f({\bar v}_{\rm ii},{\bar v}_{\so}){g(z,v_{\rm i})g(v_{\rm i},u_{\rm i})}
\mathbb{B}_1(\{{\bar u}_{\so},z\};\{{\bar v}_{\so},z\})\mathbb{B}_2(\{{\bar u}_{\rm ii},z\};\{{\bar v}_{\rm ii},z\}),
\\
%\label{3-type-3}
C_{24}=\sum_{\substack{{\bar u}\Rightarrow\{u_{\rm i},{\bar u}_{\rm ii},{\bar u}_{\st}\}\\{\bar v}\Rightarrow\{{\bar v}_{\so},v_{\rm i},{\bar v}_{\rm ii}\}} }
r_{1}^{(2)}(u_{\rm i})r_{1}^{(2)}({\bar u}_{\rm ii}) r_{3}^{(1)}(v_{\rm i})r_{3}^{(1)}({\bar v}_{\rm ii})
\frac{f({\bar u}_{\st},u_{\rm i})f({\bar u}_{\st},{\bar u}_{\rm ii})f(u_{\rm i},{\bar u}_{\rm ii})}
{f({\bar v}_{\rm ii},{\bar u}_{\rm ii})f({\bar v}_{\rm ii},u_{\rm i})f(v_{\rm i},{\bar u}_{\rm ii})f(v_{\rm i},u_{\rm i})} \\
\hphantom{C_{24}=}{}
\times f({\bar v}_{\rm ii},{\bar v}_{\so})f(v_{\rm i},{\bar v}_{\so})f({\bar v}_{\rm ii},v_{\rm i})g(z,v_{\rm i})g(u_{\rm i},z)
\mathbb{B}_1(\{{\bar u}_{\rm ii},z\};\{{\bar v}_{\so},z\})\mathbb{B}_2(\{{\bar u}_{\st},z\};\{{\bar v}_{\rm ii},z\}),
\\
%\label{3-type-2}
C_{33}=\sum_{\substack{{\bar u}\Rightarrow\{u_{\rm i},{\bar u}_{\rm ii},{\bar u}_{\st}\}\\{\bar v}\Rightarrow\{v_{\rm i},{\bar v}_{\rm ii},{\bar v}_{\st}\}} }
r_{1}^{(2)}(u_{\rm i})r_{1}^{(2)}({\bar u}_{\rm ii}) r_{3}^{(1)}({\bar v}_{\st})r_3^{(1)}(v_{\rm i})
\frac{f({\bar u}_{\st},u_{\rm i})f({\bar u}_{\st},{\bar u}_{\rm ii})f(u_{\rm i},{\bar u}_{\rm ii})}
{f({\bar v}_{\st},u_{\rm i})f({\bar v}_{\st},{\bar u}_{\rm ii})f(v_{\rm i},{\bar u}_{\rm ii})f(v_{\rm i},u_{\rm i})}\\
\hphantom{C_{33}=}{}
\times f(v_{\rm i},{\bar v}_{\rm ii})f({\bar v}_{\st},{\bar v}_{\rm ii})f({\bar v}_{\st},v_{\rm i})g(u_{\rm i},z)g(v_{\rm i},u_{\rm i})
\mathbb{B}_1(\{{\bar u}_{\rm ii},z\};\{{\bar v}_{\rm ii},z\})\mathbb{B}_2(\{{\bar u}_{\st},z\};\{{\bar v}_{\st},z\}).
\end{gather*}
One should again relabel the subsets in such a way that the Bethe vectors in all three terms have the same arguments. For this
we relabel in $C_{24}$  ${\bar u}_{\rm ii}\to {\bar u}_{\so}$ and then $ {\bar u}_{\st}\to {\bar u}_{\rm ii}$. In
$C_{33}$ we make the following replacements: ${\bar v}_{\rm ii}\to {\bar v}_{\so}$, ${\bar u}_{\rm ii}\to {\bar u}_{\so}$ and then $ {\bar v}_{\st}\to {\bar v}_{\rm ii}$,
$ {\bar u}_{\st}\to {\bar u}_{\rm ii}$. After the relabeling described above we obtain that $C_{13}+C_{24}+C_{33}=0$ due to identity
\begin{gather}\label{3-t-id}
g(z,v_{\rm i})g(v_{\rm i},u_{\rm i})+g(u_{\rm i},z)g(v_{\rm i},u_{\rm i})+g(z,v_{\rm i})g(u_{\rm i},z)=0.
\end{gather}
Thus, equation \eqref{act-T13} is proved.

\section[Action of $T_{12}$]{Action of $\boldsymbol{T_{12}}$}\label{T12}

The action of the operator $T_{12}(z)$ on $\mathcal{B}_{a,b}({\bar u};{\bar v})$ can be studied in the similar manner, but it is more
cumbersome. Therefore in this section we describe only the main steps, while the details are given in Appendix~\ref{Vanish}.

Let again ${\bar\eta}=\{z,{\bar u}\}$, ${\bar\xi}=\{z,{\bar v}\}$ with cardinalities  $\#{\bar u}=a-1$ and $\#{\bar v}=b$.
Consider a linear combination
\begin{gather}\label{lin-comb}
D=\sum g(z,\xi_0)\frac{f({\bar\xi}_0,\xi_0)}{f(z,\xi_0)}\mathcal{B}_{a,b}({\bar\eta};{\bar\xi}_0),
\end{gather}
where the sum is taken over partitions ${\bar\xi}\Rightarrow\{\xi_0,{\bar\xi}_0\}$ with $\#\xi_0=1$. It is easy to see
that this linear combination coincides with the r.h.s.\ of~\eqref{act-T12}. Indeed, at $\xi_0=z$~\eqref{lin-comb}
yields the f\/irst term in~\eqref{act-T12}, and if $\xi_0=v_0$, then~\eqref{lin-comb} yields the second term.
Due to the def\/inition $\mathcal{B}_{a,b}({\bar\eta};{\bar\xi}_0)$
we have
\begin{gather}\label{LC-BV}
D=\sum r_{1}^{(2)}({\bar\eta}_{\so}) r_{3}^{(1)}({\bar\xi}_{\st}) g(z,\xi_0)
\frac{f({\bar\eta}_{\st},{\bar\eta}_{\so})f({\bar\xi}_{\st},{\bar\xi}_{\so})f({\bar\xi}_{\so},\xi_0)f({\bar\xi}_{\st},\xi_0)}
{f({\bar\xi}_{\st},{\bar\eta}_{\so})f(z,\xi_0)}
\mathbb{B}^{(1)}({\bar\eta}_{\so};{\bar\xi}_{\so}) \mathbb{B}^{(2)}({\bar\eta}_{\st};{\bar\xi}_{\st}),\!\!\!\!
\end{gather}
where the sum is taken over partitions ${\bar\eta}\Rightarrow\{{\bar\eta}_{\so},{\bar\eta}_{\st}\}$ and ${\bar\xi}\Rightarrow\{\xi_0,{\bar\xi}_{\so},{\bar\xi}_{\st}\}$
with $\#\xi_0=1$. As before, we do not write the subscripts for the Bethe vectors~$\mathbb{B}^{(\ell)}$.
Consider again how the parameter $z$ may enter the subsets in~\eqref{LC-BV}. This time, there are f\/ive cases:
\begin{alignat*}{7}
& {\rm (i)}\quad &&{\bar\eta}_{\so}=\{z,{\bar u}_{\so}\}, \qquad && {\bar\eta}_{\st}={\bar u}_{\st}, \qquad  && {\bar\xi}_{\so}={\bar v}_{\so},  \qquad && {\bar\xi}_{\st}={\bar v}_{\st},
 \qquad && \xi_0=z,& \\
& {\rm (ii)}\quad && {\bar\eta}_{\so}=\{z,{\bar u}_{\so}\},  \qquad && {\bar\eta}_{\st}={\bar u}_{\st},  \qquad && {\bar\xi}_{\so}=\{z,{\bar v}_{\so}\},  \qquad && {\bar\xi}_{\st}={\bar v}_{\st},
 \qquad && \xi_0=v_0,& \\
& {\rm (iii)}\quad && {\bar\eta}_{\so}={\bar u}_{\so}, \qquad && {\bar\eta}_{\st}=\{z,{\bar u}_{\st}\}, \qquad && {\bar\xi}_{\so}={\bar v}_{\so},  \qquad && {\bar\xi}_{\st}={\bar v}_{\st},
 \qquad && \xi_0=z,& \\
& {\rm (iv)}\quad && {\bar\eta}_{\so}={\bar u}_{\so}, \qquad && {\bar\eta}_{\st}=\{z,{\bar u}_{\st}\}, \qquad && {\bar\xi}_{\so}=\{z,{\bar v}_{\so}\},  \qquad && {\bar\xi}_{\st}={\bar v}_{\st},
 \qquad && \xi_0=v_0,&\\
& {\rm (v)}\quad && {\bar\eta}_{\so}={\bar u}_{\so}, \qquad && {\bar\eta}_{\st}=\{z,{\bar u}_{\st}\},\qquad && {\bar\xi}_{\so}={\bar v}_{\so},\qquad && {\bar\xi}_{\st}=\{z,{\bar v}_{\st}\},
\qquad && \xi_0=v_0.& %\label{cases-1}
\end{alignat*}
In all these f\/ive cases we have dif\/ferent contributions to the linear combination $D$. So, we have
\begin{gather}\label{D-D5}
D=\sum_{\substack{{\bar u}\Rightarrow\{{\bar u}_{\so},{\bar u}_{\st}\}\\{\bar v}\Rightarrow\{{\bar v}_{\so},{\bar v}_{\st}\}} } G_{\so,\st}(D_1+D_3)
+\sum_{\substack{{\bar u}\Rightarrow\{{\bar u}_{\so},{\bar u}_{\st}\}\\{\bar v}\Rightarrow\{v_0,{\bar v}_{\so},{\bar v}_{\st}\}} } G_{0,\so,\st}(D_2+D_4+D_5),
\end{gather}
where
\begin{gather*}
%\label{G12}
G_{\so,\st} = r_{1}^{(2)}({\bar u}_{\so}) r_{3}^{(1)}({\bar v}_{\st})\frac{f({\bar u}_{\st},{\bar u}_{\so})f({\bar v}_{\st},{\bar v}_{\so})}{f({\bar v}_{\st},{\bar u}_{\so})},
\\
%\label{G012}
G_{0,\so,\st} = r_{1}^{(2)}({\bar u}_{\so}) r_{3}^{(1)}({\bar v}_{\st})
g(z,v_0) f({\bar v}_{\st},v_0)f({\bar v}_{\so},v_0) \frac{f({\bar u}_{\st},{\bar u}_{\so})f({\bar v}_{\st},{\bar v}_{\so})}{f({\bar v}_{\st},{\bar u}_{\so})},
\end{gather*}
and
\begin{gather}
\label{D1}
D_1  =   r_{1}^{(2)}(z)f({\bar u}_{\st},z)f({\bar v}_{\so},z)
\mathbb{B}^{(1)}(\{{\bar u}_{\so},z\};{\bar v}_{\so}) \mathbb{B}^{(2)}({\bar u}_{\st};{\bar v}_{\st}),
\\
\label{D2}
D_2 =  r_{1}^{(2)}(z)
f({\bar u}_{\st},z) \mathbb{B}^{(1)}(\{{\bar u}_{\so},z\};\{{\bar v}_{\so},z\})  \mathbb{B}^{(2)}({\bar u}_{\st};{\bar v}_{\st}),
\\
\label{D3}
D_3 = (z,{\bar u}_{\so})f({\bar v},z)
\mathbb{B}^{(1)}({\bar u}_{\so};{\bar v}_{\so}) \mathbb{B}^{(2)}(\{{\bar u}_{\st},z\};{\bar v}_{\st}),
\\
\label{D4}
D_4 =
f(z,{\bar u}_{\so})f({\bar v}_{\st},z)
\mathbb{B}^{(1)}({\bar u}_{\so};\{{\bar v}_{\so},z\}) \mathbb{B}^{(2)}(\{{\bar u}_{\st},z\};{\bar v}_{\st}),
\\
\label{D5}
D_5 = r_{3}^{(1)}(z)
f(z,{\bar v}_{\so}) \mathbb{B}^{(1)}({\bar u}_{\so};{\bar v}_{\so}) \mathbb{B}^{(2)}(\{{\bar u}_{\st},z\};\{{\bar v}_{\st},z\}).
\end{gather}
In all these formulas the set ${\bar u}$ is divided into subsets $\{{\bar u}_{\so},{\bar u}_{\st}\}$. As for the set ${\bar v}$, it is divided either
into subsets $\{{\bar v}_{\so},{\bar v}_{\st}\}$ (like in~\eqref{D1},~\eqref{D3}), or into subsets $\{v_0,{\bar v}_{\so},{\bar v}_{\st}\}$
with $\#v_0=1$ (like in~\eqref{D2}, \eqref{D4}, \eqref{D5}).

We should compare these contributions with the l.h.s.\ of~\eqref{act-T12}. There we have
\begin{gather*}%\label{act-T12NBV}
\frac{T_{12}(z)}{\lambda_2(z)}\mathcal{B}_{a-1,b}({\bar u};{\bar v})  =\left(\frac{T_{11}^{(2)}(z)T_{12}^{(1)}(z)}{\lambda_2^{(2)}(z)\lambda_2^{(1)}(z)}
+\frac{T_{12}^{(2)}(z)T_{22}^{(1)}(z)}{\lambda_2^{(2)}(z)\lambda_2^{(1)}(z)}
+\frac{T_{13}^{(2)}(z)T_{32}^{(1)}(z)}{\lambda_2^{(2)}(z)\lambda_2^{(1)}(z)}\right)
 \mathcal{B}_{a-1,b}({\bar u};{\bar v}).
\end{gather*}
Thus, we obtain
\begin{gather}\label{act-T12NBV-3t}
\frac{T_{12}(z)}{\lambda_2(z)}\mathcal{B}_{a-1,b}({\bar u};{\bar v})=E_1+E_2+E_3,
\end{gather}
where for $k=1,2,3$
\begin{gather*}%\label{E1}
E_k=\sum_{\substack{{\bar u}\Rightarrow\{{\bar u}_{\so},{\bar u}_{\st}\}\\{\bar v}\Rightarrow\{{\bar v}_{\so},{\bar v}_{\st}\}} }
r_{1}^{(2)}({\bar u}_{\so}) r_{3}^{(1)}({\bar v}_{\st})\frac{f({\bar u}_{\st},{\bar u}_{\so})f({\bar v}_{\st},{\bar v}_{\so})}{f({\bar v}_{\st},{\bar u}_{\so})}
\frac{T_{k2}^{(1)}(z)}{\lambda_2^{(1)}(z)}\mathbb{B}^{(1)}({\bar u}_{\so};{\bar v}_{\so})
\frac{T_{1k}^{(2)}(z)}{\lambda_2^{(2)}(z)}\mathbb{B}^{(2)}({\bar u}_{\st};{\bar v}_{\st}).
\end{gather*}
Using again the formulas of Appendix~\ref{Action} we obtain explicit expressions for the terms~$E_k$. They
are given in Appendix~\ref{Vanish}. Then similarly to the case considered in the previous section we should compare coef\/f\/icients of the products of
the partial Bethe vectors of dif\/ferent type (depending on how the parameter $z$ enters the arguments of Bethe vectors). One can show that
f\/ive terms in the combination $E_1+E_2+E_3$ reproduce the sum~\eqref{D-D5}, while other contributions mutually cancel each other. In this way
we arrive at equation~\eqref{act-T12}.

As we have explained in Section~\ref{SS-MS} the actions~\eqref{act-T13} and~\eqref{act-T12} imply the recursion~\eqref{recursion}. Since this recursion
coincides with the one for the Bethe vectors~\cite{BPRS-BV-12}, we conclude that
$\mathcal{B}_{a,b}({\bar u};{\bar v})$ \eqref{New-BV} is the Bethe vector $\mathbb{B}_{a,b}({\bar u};{\bar v})$.

\section[Coproduct property for Bethe vectors in ${\rm GL}(N)$-based models]{Coproduct property for Bethe vectors\\ in $\boldsymbol{{\rm GL}(N)}$-based models}\label{GLN}

In this section we use the results obtained in \cite{BeKhP07,KhP08}  adapted to the case under consideration.

Let $\Pi$ be a set $\{1,\ldots,N-1\}$ of indices of  simple
positive roots of $\mathfrak{gl}_N$. Let
$I=\{i_1,\dots,i_n\}$ be a f\/inite collection of positive integers.
We attach to each multiset $I$  an
ordered set of variables $\bar t_I=\{t_i\,|\,i\in I\}=\{t_{i_1},\dots,t_{i_n}\}$  with
a `coloring' map $\iota\colon \bar t_I\to\Pi$. For any multiset $I$ we assume a~natural linear
ordering in the set of variables $\bar t_I$
\begin{gather*}%\label{order}
t_{i_1}\prec\cdots\prec t_{i_n} .
\end{gather*}
Each variable $t_{i_k}$ has its own `type': $\iota(t_{i_k})\in\Pi$. We call such set of variables~$\bar t_I$ an {\it ordered
 $\Pi$-multiset}.

For example, for the ${\rm GL}(3)$-based models, the  multiset
\begin{gather}\label{setI}
I=\{1,\ldots,a,1,\ldots,b\}
\end{gather}
is used to parameterize the sets ${\bar u}$ and ${\bar v}$ of the
Bethe parameters for the Bethe vectors $\mathbb{B}_{a,b}({\bar u};{\bar v})$
\begin{gather*}%\label{BPar}
{\bar u}=\{u_1,\ldots,u_a\},\qquad {\bar v}=\{v_1,\ldots,v_b\}
\end{gather*}
and the corresponding ordering in the set $\bar t_I=\{\bar u,\bar v\}$ is
\begin{gather}\label{ord1}
u_1\prec\cdots\prec u_a\prec v_1\prec\cdots\prec v_b .
\end{gather}

Given elements $t_i,t_j$ of some ordered $\Pi$-multiset def\/ine two functions
${\gamma}(t_i,t_j)$ and $\beta(t_i,t_j)$ by the formulas
\begin{gather*}
\gamma(t_i,t_j)=
\begin{cases}
 f(t_i,t_j)^{-1} ,&  \mbox{if}\quad \iota(t_i)=\iota(t_j)+1 ,\\
 f(t_j,t_i) ,& \mbox{if}\quad \iota(t_j)=\iota(t_i)+1  ,\\
1 ,& \mbox{otherwise},
\end{cases}\qquad
%\end{gather*}
%and
%\begin{gather*}%\label{beta}
\beta(t_i,t_j)=
\begin{cases}
 f(t_j,t_i) ,& \mbox{if}\quad \iota(t_i)=\iota(t_j) ,\\
1  ,& \mbox{otherwise}.
\end{cases}
\end{gather*}

Let $V=V^{(1)}\otimes V^{(2)}$ be a tensor product of two
representations with vacuum vectors $|0\rangle^{(1)}$, $|0\rangle^{(2)}$ and weight series
$\{\lambda_b^{(1)}(u)\}$ and $\{\lambda_b^{(2)}(u)\}$,
$b=1,\ldots,N$.

A collection of rational $V$-valued functions ${\mathbf{w}}_{V,I}(t_{i}|_{i\in
I})$, depending on the representation $V$  with a
weight vacuum vector $|0\rangle$, and an ordered $\Pi$-multiset $\bar t_I$, is
called a~{\it modified weight function}~${\mathbf{w}}$ (it is the same as Bethe vectors up to normalization), if
 the following relations are satisf\/ied
\begin{itemize}\itemsep=0pt
\item ${\mathbf{w}}_{V,\varnothing}\equiv |0\rangle = |0\rangle_1\otimes |0\rangle_2$
\item  and
\begin{gather}
{\mathbf{w}}_{V,I}(t_{i} |_{i\in I})
= \sum\limits_{I\Rightarrow \{I_1,I_2\}}
{\mathbf{w}}_{V_1,I_1}(t_{i}|_{i\in I_1})
\otimes
{\mathbf{w}}_{V_2,I_2}(t_{i}|_{i\in I_2})
\cdot{\Phi}_{I_1,I_2}(t_{i}|_{i\in I}) \nonumber\\
\hphantom{{\mathbf{w}}_{V,I}(t_{i} |_{i\in I})=}{}
\times \prod\limits_{{j\in I_1}}
{\lambda^{(2)}_{\iota(t_j)}(t_{j})}
\prod\limits_{{j\in I_2}}
{\lambda^{(1)}_{\iota(t_j)+1}(t_{j})},\label{weight22}
\end{gather}
where
\begin{gather*}%\label{Phi}
{\Phi}_{I_1,I_2}(t_{i}|_{i\in I})= \prod_{i\in I_1,\ j\in I_2}
\beta(t_i,t_j) \prod_{{i\in I_2,\ j\in I_1,\ t_i\prec t_j}}\gamma(t_i,t_j) .
\end{gather*}
\end{itemize}
The summation in~(\ref{weight22}) runs over all possible decompositions of
the multiset~$I$ into a disjoint union of two non-intersecting
 multisubsets $I\Rightarrow \{I_1,I_2\}$.
The structure of ordered $\Pi$-multiset on each multisubset
is induced from that of~$I$.

For example, in the ${\rm GL}(3)$-based model the decomposition of the multiset~$I$~\eqref{setI} into two non-intersecting multisubsets~$I_1$ and~$I_2$ corresponds to the decompositions of the sets $\bar u$ and $\bar v$ into two pairs of the subsets
\begin{gather}\label{uv-sub}
{\bar u}\Rightarrow\{{\bar u}_{\so},{\bar u}_{\st}\}\qquad\mbox{and}\qquad {\bar v}\Rightarrow\{{\bar v}_{\so},{\bar v}_{\st}\}
\end{gather}
with the mutual ordering
\begin{gather*}%\label{ord2}
{\bar u}_{\so}\prec {\bar v}_{\so} ,\qquad {\bar u}_{\so}\prec {\bar v}_{\st} ,\qquad {\bar u}_{\st}\prec {\bar v}_{\so} ,\qquad
{\bar u}_{\st}\prec {\bar v}_{\st} .
\end{gather*}
inherited  by the ordering \eqref{ord1}. So the multiset $I_1$ is such that $\bar t_{I_1}=\{{\bar u}_{\so},{\bar v}_{\so}\}$ and
the corresponding to the multiset~$I_2$ is a  subset of variables $\bar t_{I_2}=\{{\bar u}_{\st},{\bar v}_{\st}\}$.

The function  ${\Phi}_{I_1,I_2}(t_{i}|_{i\in I})$ in this case is
\begin{gather*}%\label{phi1}
{\Phi}({\bar u}_{\so},{\bar u}_{\st}, {\bar v}_{\so}, {\bar v}_{\st})=f({\bar u}_{\st},{\bar u}_{\so})f({\bar v}_{\st},{\bar v}_{\so})
f({\bar v}_{\so},{\bar u}_{\st}),
\end{gather*}
and the formula \eqref{weight22} becomes{\samepage
\begin{gather}
{\mathbf{w}}_{V}({\bar u};{\bar v})
=\sum_{}
{\mathbf{w}}_{V_1}({\bar u}_{\so};{\bar v}_{\so})
\otimes
{\mathbf{w}}_{V_2}({\bar u}_{\st};{\bar v}_{\st})
\nonumber\\
\hphantom{{\mathbf{w}}_{V}({\bar u};{\bar v})=}{}
\times f({\bar u}_{\st},{\bar u}_{\so})f({\bar v}_{\st},{\bar v}_{\so}) f({\bar v}_{\so},{\bar u}_{\st})
\lambda^{(2)}_{1}({\bar u}_{\so})
\lambda^{(2)}_{2}({\bar v}_{\so})
\lambda^{(1)}_{2}({\bar u}_{\st})
\lambda^{(1)}_{3}({\bar u}_{\st}) ,\label{w22bis}
\end{gather}
where the sum is taken over the partitions~\eqref{uv-sub}.}

Def\/ine now the normalization of the  Bethe vectors that we used for the ${\rm GL}(3)$-based models  in~\cite{BPRS-BV-12}
\begin{gather}\label{normal}
\mathbb{B}_{a,b}({\bar u};{\bar v})=\frac{{\mathbf{w}}_V({\bar u};{\bar v})}{f({\bar v},{\bar u})\lambda_2({\bar u})\lambda_2({\bar v})} .
\end{gather}
Then, taking into account the coproduct property of the diagonal monodromy matrix ele\-ments~\eqref{lr}
\begin{gather*}
\lambda_2({\bar u}) =\lambda^{(1)}_2({\bar u})\lambda^{(2)}_2({\bar u})=
\lambda^{(1)}_2({\bar u}_{\so})\lambda^{(1)}_2({\bar u}_{\st})\lambda^{(2)}_2({\bar u}_{\so})\lambda^{(2)}_2({\bar u}_{\st}) ,\\
\lambda_2({\bar v}) =\lambda^{(1)}_2({\bar v})\lambda^{(2)}_2({\bar v})=
\lambda^{(1)}_2({\bar v}_{\so})\lambda^{(1)}_2({\bar v}_{\st})\lambda^{(2)}_2({\bar v}_{\so})\lambda^{(2)}_2({\bar v}_{\st}) ,
\end{gather*}
we recast \eqref{w22bis} as
\begin{gather*}%\label{w22bis2}
 \frac{{\mathbf{w}}_{V}({\bar u};{\bar v})}{f({\bar v},{\bar u})\lambda_2({\bar u})\lambda_2({\bar v})}
= \sum
\frac{{\mathbf{w}}_{V_1}({\bar u}_{\so};{\bar v}_{\so})}{f({\bar v}_{\so},{\bar u}_{\so})\lambda^{(1)}_2({\bar u}_{\so})\lambda^{(1)}_2({\bar v}_{\so})}
\otimes
\frac{{\mathbf{w}}_{V_2}({\bar u}_{\st};{\bar v}_{\st})}{f({\bar v}_{\st},{\bar u}_{\st})\lambda^{(2)}_2({\bar u}_{\st})\lambda^{(2)}_2({\bar v}_{\st})}
\\
\hphantom{\frac{{\mathbf{w}}_{V}({\bar u};{\bar v})}{f({\bar v},{\bar u})\lambda_2({\bar u})\lambda_2({\bar v})}=}{}
\times \frac{f({\bar u}_{\st},{\bar u}_{\so})f({\bar v}_{\st},{\bar v}_{\so})}{ f({\bar v}_{\st},{\bar u}_{\so})}\cdot
r^{(2)}_{1}({\bar u}_{\so}) r^{(1)}_{3}({\bar v}_{\st}) .
\end{gather*}
Under identif\/ication \eqref{normal} this relation coincides  with the statement of the Theorem~\ref{main-theor}.

\section{Conclusion}

A composite model allows one to apply the algebraic Bethe ansatz to the study of operators depending on an internal point of the
original interval. In particular, one can compute form factors of these operators and then  use the results obtained for calculating correlation functions.
A formula for the Bethe vector is a necessary tool of this program.

In this paper we  obtained an explicit expression for the Bethe vectors in the composite  quantum integrable model
associated with ${\rm GL}(3)$-invariant $R$-matrix~\eqref{R-mat}. We did it in two ways. The f\/irst one is based on the action of the monodromy matrix elements onto Bethe vectors obtained in~\cite{BPRS-BV-12}. The second way is based on the coproduct property of the weight functions proved by the completely dif\/ferent methods  in~\cite{TV95}.  Thus, we have shown that the formula that relates  Bethe vector of the composite  model with the partial Bethe vectors
is equivalent to the coproduct property.

The next step of the program is to use the formulas for the Bethe vectors in the composite  model for calculating
form factors of the entries of the partial monodromy matrices $T_{ij}^{(\ell)}$.  This will be the subject of our further publication~\cite{PRS-FFLO}.
There we consider form factors of zero modes~\cite{PakRS15a} of the operators $T_{ij}^{(\ell)}$. We show that they can be reduced to the
form factors of total monodromy matrix entries~$T_{ij}$. In this way we obtain determinant representations for form factors of local operators
in ${\rm GL}(3)$-invariant models.

\appendix

\section{Formulas of the monodromy matrix elements action\\ onto Bethe vectors}\label{Action}

We give here a list of formulas of the action of $T_{ij}(z)$ on Bethe vectors.
 These formulas were obtained in the paper \cite{BPRS-BV-12} for the more general situation of  multiple action
of the monodromy matrix elements onto Bethe vectors. In~\cite{BPRS-BV-12} these formulas were proved by induction and using
the explicit  formulas for the Bethe vectors~\eqref{BV-explicit}.

Formulas for the  actions of the partial
$T_{ij}^{(\ell)}(z)$ on partial Bethe vectors $\mathbb{B}^{(\ell)}_{a,b}$ are the same up to the replacements of $\lambda_2$ and $r_k$ by
$\lambda_2^{(\ell)}$ and $r_k^{(\ell)}$ respectively.  All formulas (except~\eqref{A-13}) are given in terms of sums over partitions
 of the original sets of the Bethe parameters  into subsets. In all the formulas $\#u_0=\#v_0=1$, and we recall that ${\bar u}_0={\bar u}\setminus u_0$
 and ${\bar v}_0={\bar v}\setminus v_0$.
\begin{itemize}\itemsep=0pt
\item Action of $T_{13}(z)$:
\begin{gather}\label{A-13}
\frac{T_{13}(z)}{\lambda_2(z)}\mathbb{B}_{a,b}({\bar u};{\bar v})=\mathbb{B}_{a+1,b+1}(\{{\bar u},z\};\{{\bar v},z\}).
\end{gather}
\item Action of $T_{12}(z)$:
\begin{gather}
\frac{T_{12}(z)}{\lambda_2(z)}\mathbb{B}_{a,b}({\bar u};{\bar v})=f({\bar v},z)\mathbb{B}_{a+1,b}(\{{\bar u},z\};{\bar v})\nonumber\\
\hphantom{\frac{T_{12}(z)}{\lambda_2(z)}\mathbb{B}_{a,b}({\bar u};{\bar v})=}{}
+\sum_{{\bar v}\Rightarrow\{v_0,{\bar v}_0\}}g(z,v_0)f({\bar v}_0,v_0)\mathbb{B}_{a+1,b}(\{{\bar u},z\};\{{\bar v}_0,z\}).\label{A-12}
\end{gather}
\item Action of $T_{23}(z)$:
\begin{gather}
\frac{T_{23}(z)}{\lambda_2(z)}\mathbb{B}_{a,b}({\bar u};{\bar v}) =f(z,{\bar u})\mathbb{B}_{a,b+1}({\bar u};\{{\bar v},z\})\nonumber\\
\hphantom{\frac{T_{23}(z)}{\lambda_2(z)}\mathbb{B}_{a,b}({\bar u};{\bar v}) =}{}
 +\sum_{{\bar u}\Rightarrow\{u_0,{\bar u}_0\}}g(u_0,z)f(u_0,{\bar u}_0)\mathbb{B}_{a,b+1}(\{{\bar u}_0,z\};\{{\bar v},z\}).\label{A-23}
\end{gather}
\item Action of $T_{11}(z)$:
\begin{gather}
\frac{T_{11}(z)}{\lambda_2(z)}\mathbb{B}_{a,b}({\bar u};{\bar v})=r_1(z)f({\bar u},z)\mathbb{B}_{a,b}({\bar u};{\bar v})\nonumber\\
\qquad{}+f({\bar v},z)\sum_{{\bar u}\Rightarrow\{u_0,{\bar u}_0\}}r_1(u_0)g(z,u_0)\frac{f({\bar u}_0,u_0)}{f({\bar v}, u_0)}\mathbb{B}_{a,b}(\{{\bar u}_0,z\};{\bar v})\nonumber\\
\qquad{}
 +\sum_{\substack{{\bar u}\Rightarrow\{u_0,{\bar u}_0\}\\{\bar v}\Rightarrow\{v_0,{\bar v}_0\}}} r_1(u_0)
g(z,v_0)g(v_0,u_0)\frac{f({\bar u}_0,u_0)f({\bar v}_0,v_0)}{f({\bar v}, u_0)}\mathbb{B}_{a,b}(\{{\bar u}_0,z\};\{{\bar v}_0,z\}).\label{A-11}
\end{gather}

\item Action of $T_{22}(z)$:
\begin{gather}
 \frac{T_{22}(z)}{\lambda_2(z)}\mathbb{B}_{a,b}({\bar u};{\bar v})\ =\  f({\bar v},z)f(z,{\bar u})\mathbb{B}_{a,b}({\bar u};{\bar v})
\nonumber\\
 \qquad{} +f(z,{\bar u})\sum_{{\bar v}\Rightarrow\{v_0,{\bar v}_0\}}g(z,v_0)f({\bar v}_0,v_0)\mathbb{B}_{a,b}({\bar u};\{{\bar v}_0,z\})
\nonumber\\
\qquad{} +f({\bar v},z)\sum_{{\bar u}\Rightarrow\{u_0,{\bar u}_0\}}g(u_0,z)f(u_0,{\bar u}_0)\mathbb{B}_{a,b}(\{{\bar u}_0,z\};{\bar v})
\nonumber\\
\qquad{} +\sum_{\substack{{\bar u}\Rightarrow\{u_0,{\bar u}_0\}\\{\bar v}\Rightarrow\{v_0,{\bar v}_0\}}}
g(z,v_0)g(u_0,z)f(u_0,{\bar u}_0)f({\bar v}_0,v_0)\mathbb{B}_{a,b}(\{{\bar u}_0,z\};\{{\bar v}_0,z\}).
\label{A-22}
\end{gather}

\item Action of $T_{33}(z)$:
\begin{gather}
\frac{T_{33}(z)}{\lambda_2(z)}\mathbb{B}_{a,b}({\bar u};{\bar v})=r_3(z)f(z,{\bar v})\mathbb{B}_{a,b}({\bar u};{\bar v})\nonumber\\
\qquad{} +f(z,{\bar u})\sum_{{\bar v}\Rightarrow\{v_0,{\bar v}_0\}}r_3(v_0)g(v_0,z)\frac{f(v_0,{\bar v}_0)}{f(v_0,{\bar u})}\mathbb{B}_{a,b}({\bar u};\{{\bar v}_0,z\})\nonumber\\
\qquad {}+\sum_{\substack{{\bar u}\Rightarrow\{u_0,{\bar u}_0\}\\{\bar v}\Rightarrow\{v_0,{\bar v}_0\}}}r_3(v_0)
g(u_0,z)g(v_0,u_0)\frac{f(u_0,{\bar u}_0)f(v_0,{\bar v}_0)}{f(v_0,{\bar u})}\mathbb{B}_{a,b}(\{{\bar u}_0,z\};\{{\bar v}_0,z\}).\label{A-33}
\end{gather}

\item Action of $T_{32}(z)$:
\begin{gather}
\frac{T_{32}(z)}{\lambda_2(z)}\mathbb{B}_{a,b}({\bar u};{\bar v}) \nonumber\\
\qquad{}=\sum_{\substack{{\bar u}\Rightarrow\{u_0,{\bar u}_0\}\\{\bar v}\Rightarrow\{v_0,{\bar v}_0\}}}
r_3(v_0)g(u_0,z)g(v_0,u_0)\frac{f(u_0,{\bar u}_0)f(v_0,{\bar v}_0) f({\bar v}_0,z)}{f(v_0,{\bar u})}
\mathbb{B}_{a,b-1}(\{{\bar u}_0,z\};{\bar v}_0)\nonumber\\
\qquad\quad{} +\sum_{{\bar v}\Rightarrow\{v_0,{\bar v}_0\}}g(z,v_0)\Bigl(r_3(z)f(z,{\bar v}_0)f({\bar v}_0,v_0)-r_3(v_0)\frac{f({\bar v}_0,z)f(v_0,{\bar v}_0)f(z,{\bar u})}
{f(v_0,{\bar u})}\Bigr)\nonumber\\
\qquad\qquad{}\times \mathbb{B}_{a,b-1}({\bar u};{\bar v}_0)\nonumber\\
\qquad{} +\sum_{{\bar v}\Rightarrow\{v_0,v_1,{\bar v}_2\}}r_3(v_0)g(v_0,z)g(z,v_1)\frac{f(v_0,v_1)f(v_0,{\bar v}_2)f({\bar v}_2,v_1)f(z,{\bar u})}
{f(v_0,{\bar u})}\nonumber\\
 \qquad\qquad{}\times
\mathbb{B}_{a,b-1}({\bar u};\{{\bar v}_2,z\})\nonumber\\
\qquad\quad{} +\sum_{\substack{{\bar u}\Rightarrow\{u_0,{\bar u}_0\}\\{\bar v}\Rightarrow\{v_0,v_1{\bar v}_2\}}}
r_3(v_0)g(u_0,z)g(z,v_1)g(v_0,u_0)\frac{f(v_0,v_1)f(v_0,{\bar v}_2)f({\bar v}_2,v_1)f(u_0,{\bar u}_0)}
{f(v_0,{\bar u})}\nonumber\\
\qquad\qquad{}
\times\mathbb{B}_{a,b-1}(\{{\bar u}_0,z\};\{{\bar v}_2,z\}).\label{A-32}
\end{gather}
In this formula $\#v_0=\#v_1=\#u_0=1$.
\end{itemize}

\section{Proof of the equality (\ref{act-T12})} \label{Vanish}

In this appendix we give explicit formulas for the terms $E_k$ in \eqref{act-T12NBV-3t}.
It is clear from the formulas  of Appendix~\ref{Action} that
  $E_1$, $E_2$ and $E_3$ respectively consist of six, eight and f\/ive dif\/ferent sums over partitions. We denote them as
  $E_{1,m}$, $m=1,\ldots,6$, $E_{2,m}$, $m=1,\ldots,8$, and $E_{3,m}$, $m=1,\ldots,5$.
  Each of these contributions can be presented
in the form
\begin{gather}\label{G-g}
E_{k,m}  = \sum_{\substack{{\bar u}\Rightarrow\{{\bar u}_{\so},{\bar u}_{\st}\}\\{\bar v}\Rightarrow\{{\bar v}_{\so},{\bar v}_{\st}\}} }
r_{1}^{(2)}({\bar u}_{\so}) r_{3}^{(1)}({\bar v}_{\st})\frac{f({\bar u}_{\st},{\bar u}_{\so})f({\bar v}_{\st},{\bar v}_{\so})}{f({\bar v}_{\st},{\bar u}_{\so})}  \gamma_{k,m},
\end{gather}
where $\gamma_{k,m}$ may contain additional sums over partitions.

Due to \eqref{A-12} and \eqref{A-11} we have for the term $E_1$:
\begin{gather}\label{C11}
\gamma_{1,1}=r_1^{(2)}(z) f({\bar v}_{\so},z)f({\bar u}_{\st},z)  \mathbb{B}^{(1)}(\{{\bar u}_{\so},z\};{\bar v}_{\so})
\mathbb{B}^{(2)}({\bar u}_{\st};{\bar v}_{\st}),
\\
\gamma_{1,2}=\sum_{{\bar u}_{\st}\Rightarrow\{u_{\rm i},{\bar u}_{\rm ii}\}}r_1^{(2)}(u_{\rm i})g(z,u_{\rm i})
\frac{f({\bar v}_{\so},z)f({\bar v}_{\st},z)f({\bar u}_{\rm ii},u_{\rm i})}{f({\bar v}_{\st}, u_{\rm i})} \nonumber\\
\hphantom{\gamma_{1,2}=}{}\times
\mathbb{B}^{(1)}(\{{\bar u}_{\so},z\};{\bar v}_{\so})
\mathbb{B}^{(2)}(\{{\bar u}_{\rm ii},z\};{\bar v}_{\st}),\label{C12}
\\
\gamma_{1,3}=\sum_{\substack{{\bar u}_{\st}\Rightarrow\{u_{\rm i},{\bar u}_{\rm ii}\}\\{\bar v}_{\st}\Rightarrow\{v_{\rm i},{\bar v}_{\rm ii}\}}} r_1^{(2)}(u_{\rm i})
g(z,v_{\rm i})g(v_{\rm i},u_{\rm i})
\frac{f({\bar v}_{\so},z)f({\bar u}_{\rm ii},u_{\rm i})f({\bar v}_{\rm ii},v_{\rm i})}{f({\bar v}_{\st}, u_{\rm i})}\nonumber\\
\hphantom{\gamma_{1,3}=}{}
\times\mathbb{B}^{(1)}(\{{\bar u}_{\so},z\};{\bar v}_{\so})\mathbb{B}^{(2)}(\{{\bar u}_{\rm ii},z\};\{{\bar v}_{\rm ii},z\}),\label{C13}
\\
\gamma_{1,4}=\sum_{{\bar v}_{\so}\Rightarrow\{v_{\rm iii},{\bar v}_{\rm iv}\}}r_1^{(2)}(z)g(z,v_{\rm iii})f({\bar v}_{\rm iv},v_{\rm iii})
f({\bar u}_{\st},z)  \mathbb{B}^{(1)}(\{{\bar u}_{\so},z\};\{{\bar v}_{\rm iv},z\})\mathbb{B}^{(2)}({\bar u}_{\st};{\bar v}_{\st}),\label{C14}
\\
\gamma_{1,5}=\sum_{\substack{{\bar v}_{\so}\Rightarrow\{v_{\rm iii},{\bar v}_{\rm iv}\}\\ {\bar u}_{\st}\Rightarrow\{u_{\rm i},{\bar u}_{\rm ii}\}}}
r_1^{(2)}(u_{\rm i})g(z,u_{\rm i}) g(z,v_{\rm iii})
\frac{f({\bar v}_{\rm iv},v_{\rm iii})f({\bar v}_{\st},z)f({\bar u}_{\rm ii},u_{\rm i})}
{f({\bar v}_{\st}, u_{\rm i})}\nonumber\\
\hphantom{\gamma_{1,5}=}{}
\times\mathbb{B}^{(1)}(\{{\bar u}_{\so},z\};\{{\bar v}_{\rm iv},z\})\mathbb{B}^{(2)}(\{{\bar u}_{\rm ii},z\};{\bar v}_{\st}),\label{C15}
\\
\gamma_{1,6}=\sum_{\substack{{\bar v}_{\so}\Rightarrow\{v_{\rm iii},{\bar v}_{\rm iv}\}\\
\substack{{\bar u}_{\st}\Rightarrow\{u_{\rm i},{\bar u}_{\rm ii}\}\\{\bar v}_{\st}\Rightarrow\{v_{\rm i},{\bar v}_{\rm ii}\}}  }}
 r_1^{(2)}(u_{\rm i})
g(z,v_{\rm iii})g(z,v_{\rm i})g(v_{\rm i},u_{\rm i})
\frac{f({\bar v}_{\rm iv},v_{\rm iii})f({\bar u}_{\rm ii},u_{\rm i})f({\bar v}_{\rm ii},v_{\rm i})}{f({\bar v}_{\st}, u_{\rm i})}\nonumber\\
\hphantom{\gamma_{1,6}=}{}
\times\mathbb{B}^{(1)}(\{{\bar u}_{\so},z\};\{{\bar v}_{\rm iv},z\})\mathbb{B}^{(2)}(\{{\bar u}_{\rm ii},z\};\{{\bar v}_{\rm ii},z\}).\label{C16}
\end{gather}

Due to \eqref{A-22} and \eqref{A-12} we have for the term $E_2$:
\begin{gather}\label{C21}
\gamma_{2,1}=f({\bar v}_{\so},z)f(z,{\bar u}_{\so})f({\bar v}_{\st},z) \mathbb{B}^{(1)}({\bar u}_{\so};{\bar v}_{\so})\mathbb{B}^{(2)}(\{{\bar u}_{\st},z\};{\bar v}_{\st}),
\\
\label{C22}
\gamma_{2,2}=\sum_{{\bar v}_{\so}\Rightarrow\{v_{\rm iii},{\bar v}_{\rm iv}\}}
g(z,v_{\rm iii})
f(z,{\bar u}_{\so})f({\bar v}_{\rm iv},v_{\rm iii})f({\bar v}_{\st},z) \mathbb{B}^{(1)}({\bar u}_{\so};\{{\bar v}_{\rm iv},z\})\mathbb{B}^{(2)}(\{{\bar u}_{\st},z\};{\bar v}_{\st}),
\\
\label{C23}
\gamma_{2,3}=\sum_{{\bar u}_{\so}\Rightarrow\{u_{\rm i},{\bar u}_{\rm ii}\}}
g(u_{\rm i},z)f({\bar v}_{\so},z)f(u_{\rm i},{\bar u}_{\rm ii})f({\bar v}_{\st},z)
\mathbb{B}^{(1)}(\{{\bar u}_{\rm ii},z\};{\bar v}_{\so})\mathbb{B}^{(2)}(\{{\bar u}_{\st},z\};{\bar v}_{\st}),
\\
\gamma_{2,4}=\sum_{\substack{{\bar u}_{\so}\Rightarrow\{u_{\rm i},{\bar u}_{\rm ii}\}\\{\bar v}_{\so}\Rightarrow\{v_{\rm iii},{\bar v}_{\rm iv}\}}}
g(z,v_{\rm iii})g(u_{\rm i},z)f(u_{\rm i},{\bar u}_{\rm ii})f({\bar v}_{\rm iv},v_{\rm iii})f({\bar v}_{\st},z)\nonumber\\
\hphantom{\gamma_{2,4}=}{}
\times\mathbb{B}^{(1)}(\{{\bar u}_{\rm ii},z\};\{{\bar v}_{\rm iv},z\})\mathbb{B}^{(2)}(\{{\bar u}_{\st},z\};{\bar v}_{\st}),\label{C24}
\\
\label{C25}
\gamma_{2,5}=\sum_{{\bar v}_{\st}\Rightarrow\{v_{\rm i},{\bar v}_{\rm ii}\}}g(z,v_{\rm i}) f({\bar v}_{\so},z)f(z,{\bar u}_{\so})f({\bar v}_{\rm ii},v_{\rm i})
\mathbb{B}^{(1)}({\bar u}_{\so};{\bar v}_{\so})\mathbb{B}^{(2)}(\{{\bar u}_{\st},z\};\{{\bar v}_{\rm ii},z\}),
\\
\gamma_{2,6}=\sum_{\substack{{\bar v}_{\so}\Rightarrow\{v_{\rm iii},{\bar v}_{\rm iv}\}\\  {\bar v}_{\st}\Rightarrow\{v_{\rm i},{\bar v}_{\rm ii}\} }}
g(z,v_{\rm iii})g(z,v_{\rm i})
f(z,{\bar u}_{\so})f({\bar v}_{\rm iv},v_{\rm iii})f({\bar v}_{\rm ii},v_{\rm i})\nonumber\\
\hphantom{\gamma_{2,6}=}{}
\times \mathbb{B}^{(1)}({\bar u}_{\so};\{{\bar v}_{\rm iv},z\})
\mathbb{B}^{(2)}(\{{\bar u}_{\st},z\};\{{\bar v}_{\rm ii},z\}),\label{C26}
\\
\gamma_{2,7}=\sum_{\substack{{\bar u}_{\so}\Rightarrow\{u_{\rm i},{\bar u}_{\rm ii}\}\\  {\bar v}_{\st}\Rightarrow\{v_{\rm i},{\bar v}_{\rm ii}\} }}
g(u_{\rm i},z)g(z,v_{\rm i})f(u_{\rm i},{\bar u}_{\rm ii}) f({\bar v}_{\so},z)f({\bar v}_{\rm ii},v_{\rm i})\nonumber\\
\hphantom{\gamma_{2,7}=}{}
\times\mathbb{B}^{(1)}(\{{\bar u}_{\rm ii},z\};{\bar v}_{\so})
\mathbb{B}^{(2)}(\{{\bar u}_{\st},z\};\{{\bar v}_{\rm ii},z\}),\label{C27}
\\
\gamma_{2,8}=\sum_{\substack{\substack{{\bar u}_{\so}\Rightarrow\{u_{\rm i},{\bar u}_{\rm ii}\}\\{\bar v}_{\so}\Rightarrow\{v_{\rm iii},{\bar v}_{\rm iv}\}}\\
{\bar v}_{\st}\Rightarrow\{v_{\rm i},{\bar v}_{\rm ii}\} }}
g(z,v_{\rm iii})g(u_{\rm i},z)g(z,v_{\rm i})
f(u_{\rm i},{\bar u}_{\rm ii})f({\bar v}_{\rm iv},v_{\rm iii})f({\bar v}_{\rm ii},v_{\rm i})\nonumber\\
\hphantom{\gamma_{2,8}=}{}
\times \mathbb{B}^{(1)}(\{{\bar u}_{\rm ii},z\};\{{\bar v}_{\rm iv},z\})
\mathbb{B}^{(2)}(\{{\bar u}_{\st},z\};\{{\bar v}_{\rm ii},z\}).\label{C28}
\end{gather}

Due to \eqref{A-32} and \eqref{A-13} we have for the term $E_3$:
\begin{gather}
\gamma_{3,1}=\sum_{\substack{{\bar u}_{\so}\Rightarrow\{u_{\rm i},{\bar u}_{\rm ii}\}\\{\bar v}_{\so}\Rightarrow\{v_{\rm i},{\bar v}_{\rm iii}\}}}
r_{3}^{(1)}(v_{\rm i})g(u_{\rm i},z)g(v_{\rm i},u_{\rm i})\frac{f(u_{\rm i},{\bar u}_{\rm ii})f(v_{\rm i},{\bar v}_{\rm iii}) f({\bar v}_{\rm iii},z)}{f(v_{\rm i},{\bar u}_{\so})}\nonumber
\\
\hphantom{\gamma_{3,1}=}{}
\times\mathbb{B}^{(1)}(\{{\bar u}_{\rm ii},z\};{\bar v}_{\rm iii})\mathbb{B}^{(2)}(\{{\bar u}_{\st},z\};\{{\bar v}_{\st},z\}),\label{C31}
\\
\gamma_{3,2}=\sum_{{\bar v}_{\so}\Rightarrow\{v_{\rm i},{\bar v}_{\rm iii}\}}{r_{3}^{(1)}(z){g(z,v_{\rm i})}f(z,{\bar v}_{\rm iii})
f({\bar v}_{\rm iii},v_{\rm i})}
{\mathbb{B}^{(1)}({\bar u}_{\so};{\bar v}_{\rm iii})}\mathbb{B}^{(2)}(\{{\bar u}_{\st},z\};\{{\bar v}_{\st},z\}),\label{C32}
\\
\gamma_{3,3}=\sum_{{\bar v}_{\so}\Rightarrow\{v_{\rm i},{\bar v}_{\rm iii}\}}r_{3}^{(1)}(v_{\rm i}){g(v_{\rm i},z)}
\frac{f({\bar v}_{\rm iii},z)f(v_{\rm i},{\bar v}_{\rm iii})f(z,{\bar u}_{\so})}
{f(v_{\rm i},{\bar u}_{\so})}\nonumber\\
\hphantom{\gamma_{3,3}=}{}
\times{\mathbb{B}^{(1)}({\bar u}_{\so};{\bar v}_{\rm iii})}\mathbb{B}^{(2)}(\{{\bar u}_{\st},z\};\{{\bar v}_{\st},z\}),\label{C33}
\\
\gamma_{3,4}=\sum_{{\bar v}_{\so}\Rightarrow\{v_{\rm i},v_{\rm ii},{\bar v}_{\rm iii}\}}r_{3}^{(1)}(v_{\rm i})g(v_{\rm i},z)g(z,v_{\rm ii})
\frac{f(v_{\rm i},v_{\rm ii})f(v_{\rm i},{\bar v}_{\rm iii})f({\bar v}_{\rm iii},v_{\rm ii})f(z,{\bar u}_{\so})}
{f(v_{\rm i},{\bar u}_{\so})}\nonumber\\
\hphantom{\gamma_{3,4}=}{}
\times\mathbb{B}^{(1)}({\bar u}_{\so};\{{\bar v}_{\rm iii},z\})\mathbb{B}^{(2)}(\{{\bar u}_{\st},z\};\{{\bar v}_{\st},z\}),\label{C34}
\\
\gamma_{3,5}=\sum_{\substack{{\bar u}_{\so}\Rightarrow\{u_{\rm i},{\bar u}_{\rm ii}\}\\{\bar v}_{\so}\Rightarrow\{v_{\rm i},v_{\rm ii}{\bar v}_{\rm iii}\}}}
r_{3}^{(1)}(v_{\rm i})g(u_{\rm i},z)g(z,v_{\rm ii})g(v_{\rm i},u_{\rm i})\frac{f(v_{\rm i},v_{\rm ii})f(v_{\rm i},{\bar v}_{\rm iii})f({\bar v}_{\rm iii},v_{\rm ii})f(u_{\rm i},{\bar u}_{\rm ii})}
{f(v_{\rm i},{\bar u}_{\so})}\nonumber\\
\hphantom{\gamma_{3,5}=}{}
\times \mathbb{B}^{(1)}(\{{\bar u}_{\rm ii},z\};\{{\bar v}_{\rm iii},z\})\mathbb{B}^{(2)}(\{{\bar u}_{\st},z\};\{{\bar v}_{\st},z\}).\label{C35}
\end{gather}

First of all one can easily observe that contributions $\gamma_{1,1}$ \eqref{C11} and $\gamma_{2,1}$~\eqref{C21} into the sum~\eqref{act-T12NBV-3t}  coincide identically with the contributions~$D_1$~\eqref{D1} and~$D_3$~\eqref{D3} into the f\/irst sum of~\eqref{D-D5}.
The contributions  $\gamma_{1,4}$~\eqref{C14}, $\gamma_{2,2}$~\eqref{C22} and
 $\gamma_{3,2}$~\eqref{C32} coincide with the contributions
$D_2$~\eqref{D2}, $D_4$~\eqref{D4} and~$D_5$~\eqref{D5} into the second sum of~\eqref{D-D5} up to  a relabeling
of the subsets. Indeed, consider, for example, the contribution $E_{1,4}$.   Combining~\eqref{G-g} and~\eqref{C14}
we obtain
\begin{gather*}%\label{G-g14}
E_{1,4}  = \sum_{\substack{{\bar u}\Rightarrow\{{\bar u}_{\so},{\bar u}_{\st}\}\\{\bar v}\Rightarrow\{v_{\rm iii},{\bar v}_{\rm iv},{\bar v}_{\st}\}} }
r_{1}^{(2)}({\bar u}_{\so}) r_{3}^{(1)}({\bar v}_{\st})\frac{f({\bar u}_{\st},{\bar u}_{\so})f({\bar v}_{\st},v_{\rm iii})f({\bar v}_{\st},{\bar v}_{\rm iv})}
{f({\bar v}_{\st},{\bar u}_{\so})}\\
\hphantom{E_{1,4}  =}{}
\times r_1^{(2)}(z)g(z,v_{\rm iii})f({\bar v}_{\rm iv},v_{\rm iii})
f({\bar u}_{\st},z)  \mathbb{B}^{(1)}(\{{\bar u}_{\so},z\};\{{\bar v}_{\rm iv},z\})\mathbb{B}^{(2)}({\bar u}_{\st};{\bar v}_{\st}) .
\end{gather*}
Relabeling of the subsets $v_{{\rm iii}}\to v_0$ and ${\bar v}_{{\rm iv}}\to {\bar v}_{\so}$ leads to
\begin{gather*}
E_{1,4}  = \sum_{\substack{{\bar u}\Rightarrow\{{\bar u}_{\so},{\bar u}_{\st}\}\\{\bar v}\Rightarrow\{v_0,{\bar v}_{\so},{\bar v}_{\st}\}} }
r_{1}^{(2)}({\bar u}_{\so}) r_{3}^{(1)}({\bar v}_{\st})\frac{f({\bar u}_{\st},{\bar u}_{\so})f({\bar v}_{\st},v_{0})f({\bar v}_{\st},{\bar v}_{\so})}
{f({\bar v}_{\st},{\bar u}_{\so})}\\
\hphantom{E_{1,4}  =}{}
\times r_1^{(2)}(z)g(z,v_0)f({\bar v}_{\so},v_0)
f({\bar u}_{\st},z)  \mathbb{B}^{(1)}(\{{\bar u}_{\so},z\};\{{\bar v}_{\so},z\})\mathbb{B}^{(2)}({\bar u}_{\st};{\bar v}_{\st}),%\label{G-g14-2}
\end{gather*}
which exactly coincides with the sum over
partitions~\eqref{D-D5} of the contribution~$D_2$.
The identif\/ications  of the contributions~$E_{2,2}$ and
$E_{3,2}$ with the terms~$D_4$ and~$D_5$ in~\eqref{D-D5} can be  done similarly.

All other contributions should vanish. To check this we  follow the same strategy as in
Section~\ref{T13}. We consider separately the coef\/f\/icients of the products of the Bethe vectors
$\mathbb{B}^{(1)}\mathbb{B}^{(2)}$ with a dif\/ferent placement of the parameter~$z$.
There are six dif\/ferent  types of these products:
\begin{gather*}%\label{prod}
  \mathbb{B}^{(1)}(\{{\bar u}',z\};\{{\bar v}'\})\mathbb{B}^{(2)}(\{{\bar u}'',z\};\{{\bar v}''\}), \\
  \mathbb{B}^{(1)}(\{{\bar u}',z\};\{{\bar v}'\})\mathbb{B}^{(2)}(\{{\bar u}'',z\};\{{\bar v}'',z\}),\\
  \mathbb{B}^{(1)}(\{{\bar u}'\};\{{\bar v}'\})\mathbb{B}^{(2)}(\{{\bar u}'',z\};\{{\bar v}'',z\}),\\
  \mathbb{B}^{(1)}(\{{\bar u}',z\};\{{\bar v}',z\})\mathbb{B}^{(2)}(\{{\bar u}'',z\};\{{\bar v}''\}),\\
  \mathbb{B}^{(1)}(\{{\bar u}'\};\{{\bar v}',z\})\mathbb{B}^{(2)}(\{{\bar u}'',z\};\{{\bar v}'',z\}),\\
  \mathbb{B}^{(1)}(\{{\bar u}',z\};\{{\bar v}',z\})\mathbb{B}^{(2)}(\{{\bar u}'',z\};\{{\bar v}'',z\}).
\end{gather*}
\begin{itemize}\itemsep=0pt
\item Coef\/f\/icients of $\mathbb{B}^{(1)}(\{{\bar u}',z\};\{{\bar v}'\})\mathbb{B}^{(2)}(\{{\bar u}'',z\};\{{\bar v}''\})$
 come from \eqref{C12} and  \eqref{C23}.
 \item
Coef\/f\/icients of $\mathbb{B}^{(1)}(\{{\bar u}'\};\{{\bar v}'\})\mathbb{B}^{(2)}(\{{\bar u}'',z\};\{{\bar v}'',z\})$
 come from \eqref{C25} and \eqref{C33}.
 \item
 Coef\/f\/icients of $\mathbb{B}^{(1)}(\{{\bar u}',z\};\{{\bar v}',z\})\mathbb{B}^{(2)}(\{{\bar u}'',z\};\{{\bar v}''\})$
 come from \eqref{C15}  and \eqref{C24}.
 \item
Coef\/f\/icients of $\mathbb{B}^{(1)}(\{{\bar u}'\};\{{\bar v}',z\})\mathbb{B}^{(2)}(\{{\bar u}'',z\};\{{\bar v}'',z\})$
 come from \eqref{C26} and  \eqref{C34}.
 \end{itemize}

 In all these four cases the cancellations of the coef\/f\/icients after  relabeling the subsets occurs due to the
 identity~\eqref{1-t-id}.

 Coef\/f\/icients of $\mathbb{B}^{(1)}(\{{\bar u}',z\};\{{\bar v}'\})\mathbb{B}^{(2)}(\{{\bar u}'',z\};\{{\bar v}'',z\})$
 come from \eqref{C13}, \eqref{C27} and  \eqref{C31}.  After proper  relabeling  the subsets they cancel each other due to the
 identity~\eqref{3-t-id}.

Finally, the  coef\/f\/icients of $\mathbb{B}^{(1)}(\{{\bar u}',z\};\{{\bar v}',z\})\mathbb{B}^{(2)}(\{{\bar u}'',z\};\{{\bar v}'',z\})$
 come from~\eqref{C16},  \eqref{C28} and~\eqref{C35}.
 In this case again after  relabeling the subsets they vanish due to the identity~\eqref{3-t-id}.

\subsection*{Acknowledgements}

The work of S.P.~was supported in part by RFBR-Ukraine grant 14-01-90405-ukr-a.
N.A.S.~was  supported by the Program of RAS ``Nonlinear Dynamics in Mathematics and Physics''
and  by the grant RFBR-15-31-20484-mol$\_$a$\_$ved.

%\cite{BPRS13}

\pdfbookmark[1]{References}{ref}
\LastPageEnding


\begin{thebibliography}{99}
\footnotesize \itemsep=0pt

\bibitem{BPRS-BV-12}
Belliard S., Pakuliak S., Ragoucy E., Slavnov N.A., Bethe vectors of {${\rm
  GL}(3)$}-invariant integrable models, \href{http://dx.doi.org/10.1088/1742-5468/2013/02/P02020}{\textit{J.~Stat. Mech. Theory Exp.}}
  \textbf{2013} (2013), P02020, 24~pages, \href{http://arxiv.org/abs/1210.0768}{arXiv:1210.0768}.

\bibitem{BeKhP07}
Enriquez B., Khoroshkin S., Pakuliak S., Weight functions and {D}rinfeld
  currents, \href{http://dx.doi.org/10.1007/s00220-007-0351-y}{\textit{Comm. Math. Phys.}} \textbf{276} (2007), 691--725,
  \href{http://arxiv.org/abs/math.QA/0610398}{math.QA/0610398}.

\bibitem{Ize87}
Izergin A.G., Partition function of a six-vertex model in a f\/inite volume,
  \textit{Sov. Phys. Dokl.} \textbf{32} (1987), 878--879.

\bibitem{IzK84}
Izergin A.G., Korepin V.E., The quantum inverse scattering method approach to
  correlation functions, \href{http://dx.doi.org/10.1007/BF01212350}{\textit{Comm. Math. Phys.}} \textbf{94} (1984), 67--92.

\bibitem{IzKR87}
Izergin A.G., Korepin V.E., Reshetikhin N.Yu., Correlation functions in a
  one-dimensional {B}ose gas, \href{http://dx.doi.org/10.1088/0305-4470/20/14/022}{\textit{J.~Phys.~A: Math. Gen.}} \textbf{20}
  (1987), 4799--4822.

\bibitem{KhP08}
Khoroshkin S., Pakuliak S., A computation of universal weight function for
  quantum af\/f\/ine algebra {$U_q(\widehat{\mathfrak{gl}}_N)$}, \textit{J.~Math.
  Kyoto Univ.} \textbf{48} (2008), 277--321, \href{http://arxiv.org/abs/0711.2819}{arXiv:0711.2819}.

\bibitem{KhPT}
Khoroshkin S., Pakuliak S., Tarasov V., Of\/f-shell {B}ethe vectors and
  {D}rinfeld currents, \href{http://dx.doi.org/10.1016/j.geomphys.2007.02.005}{\textit{J.~Geom. Phys.}} \textbf{57} (2007), 1713--1732,
  \href{http://arxiv.org/abs/math.QA/0610517}{math.QA/0610517}.

\bibitem{KitKMST07}
Kitanine N., Kozlowski K., Maillet J.M., Slavnov N.A., Terras V., On
  correlation functions of integrable models associated with the six-vertex
  {$R$}-matrix, \href{http://dx.doi.org/10.1088/1742-5468/2007/01/P01022}{\textit{J.~Stat. Mech. Theory Exp.}} \textbf{2007} (2007),
  P01022, 17~pages, \href{http://arxiv.org/abs/hep-th/0611142}{hep-th/0611142}.

\bibitem{KMT99}
Kitanine N., Maillet J.M., Terras V., Correlation functions of the {$XXZ$}
  {H}eisenberg spin-{${1\over2}$} chain in a~magnetic f\/ield, \href{http://dx.doi.org/10.1016/S0550-3213(99)00619-7}{\textit{Nuclear
  Phys.~B}} \textbf{567} (2000), 554--582, \href{http://arxiv.org/abs/math-ph/9907019}{math-ph/9907019}.

\bibitem{Kor82}
Korepin V.E., Calculation of norms of {B}ethe wave functions, \href{http://dx.doi.org/10.1007/BF01212176}{\textit{Comm.
  Math. Phys.}} \textbf{86} (1982), 391--418.


\bibitem{PRS14}
Pakuliak S., Ragoucy E., Slavnov N.A., Bethe vectors of quantum integrable
  models based on {$U_q(\widehat{\mathfrak{gl}}_N)$}, \href{http://dx.doi.org/10.1088/1751-8113/47/10/105202}{\textit{J.~Phys.~A: Math.
  Theor.}} \textbf{47} (2014), 105202, 16~pages, \href{http://arxiv.org/abs/1310.3253}{arXiv:1310.3253}.

\bibitem{PakRS15a}
Pakuliak S., Ragoucy E., Slavnov N.A., Zero modes method and form factors in
  quantum integrable models, \href{http://dx.doi.org/10.1016/j.nuclphysb.2015.02.006}{\textit{Nuclear Phys.~B}} \textbf{893} (2015),
  459--481, \href{http://arxiv.org/abs/1412.6037}{arXiv:1412.6037}.

\bibitem{PRS-FFLO}
Pakuliak S., Ragoucy E., Slavnov N.A., Form factors of local operators in a
  one-dimensional two-component Bose gas, \href{http://arxiv.org/abs/1503.00546}{arXiv:1503.00546}.

\bibitem{Sl89}
Slavnov N.A., Calculation of scalar products of wave functions and form-factors
  in the framework of the algebraic {B}ethe ansatz, \href{http://dx.doi.org/10.1007/BF01016531}{\textit{Theoret. and Math.
  Phys.}} \textbf{79} (1989), 502--508.

\bibitem{Sl07}
Slavnov N.A., The algebraic {B}ethe ansatz and quantum integrable systems,
  \href{http://dx.doi.org/10.1070/RM2007v062n04ABEH004430}{\textit{Russian Math. Surveys}} \textbf{62} (2007), 727--766.

\bibitem{TV95}
Varchenko A.N., Tarasov V.O., Jackson integral representations for solutions of
  the {K}nizhnik--{Z}amolodchikov quantum equation, \textit{St.~Petersburg
  Math.~J.} \textbf{6} (1994), 275--313, \href{http://arxiv.org/abs/hep-th/9311040}{hep-th/9311040}.

\end{thebibliography}
\end{document}